\documentclass[aps, onecolumn, prx,superscriptaddress, notitlepage]{revtex4-1}
\usepackage{bm,amsmath,graphicx, mathtools,multirow,gensymb}
\usepackage[colorlinks=true,linkcolor=MidnightBlue,urlcolor=black,citecolor=MidnightBlue,anchorcolor=MidnightBlue]{hyperref}\usepackage[dvipsnames]{xcolor}
\begin{document}
%%%%%====================================
%%%%%====================================
\title{
% Rigid flocks and chiral foldamers d a chemically active polymer}
%%
%Rigid flocks, crystallites, and chemotactic swimmers in a chemically self-interacting active %colloidal chain
%%
Rigid flocks, undulatory gaits, and chiral foldamers
%from chemical self-interaction 
in a chemically active polymer}
% in a chemically self-interacting active colloidal chain
% }
%%
%\title{Emergent rigidity, chemotactic swimming, and chirality in a chemically active colloidal chain}
%%
\author{Arvin Gopal Subramaniam}
\email{ph22d800@smail.iitm.ac.in}
\affiliation{Department of Physics, Indian Institute of Technology Madras, Chennai, India}
\affiliation{Center for Soft and Biological Matter, IIT Madras, Chennai,  India}
\author{Manoj Kumar}
\affiliation{Simons Centre for the Study of Living Machines, National Centre for
Biological Sciences, Tata Institute of Fundamental Research, Bangalore,
India}
\author{Shashi Thutupalli}
\affiliation{Simons Centre for the Study of Living Machines, National Centre for
Biological Sciences, Tata Institute of Fundamental Research, Bangalore,
India}
\affiliation{International Centre for Theoretical Sciences, Tata Institute of Fundamental
Research, Bangalore, India}
\author{Rajesh Singh}
\email{rsingh@physics.iitm.ac.in}
\affiliation{Department of Physics, Indian Institute of Technology Madras, Chennai, India}
\affiliation{Center for Soft and Biological Matter, IIT Madras, Chennai,  India}
\begin{abstract}
Active matter systems - such as a collection of active colloidal particles - operate far from equilibrium with complex inter-particle interactions that govern their collective dynamics. Predicting the collective dynamics of such systems may aid the design of self-shaping structures comprised of active colloidal units with a prescribed dynamical function. Here, using simulations and theory, we study the collective dynamics of a  chain consisting of active Brownian particles with internal interactions via trail-mediated chemicals, connected by harmonic springs in two dimensions to obtain design principles for active colloidal molecules. We show that two-dimensional confinement and chemo-repulsive interactions between the freely-jointed particles lead to an emergent rigidity of the chain in the steady-state dynamics. In the chemo-attractive regime, the chain collapses into crystals that abruptly halt their motion. Further, in a chain consisting of a binary mixture of monomers, we show that non-reciprocal chemical affinities between distinct species give rise to novel phenomena, such as chiral molecules with tunable dynamics, sustained undulatory gaits and reversal of the direction of motion. Our results suggest a novel interpretation of the role of trail-mediated interactions, in addition to providing active self-assembly principles arising due to non-reciprocal interactions.
\end{abstract}
\maketitle
\section{Introduction}\label{sec:intro}
Active colloidal particles are a class of active matter that exhibit rich collective phenomenon at the many-body dynamical level \cite{cates2015motility, bechinger2016}. Such systems include those of biological \cite{goldstein2015green} or synthetic \cite{ebbens2010pursuit} origin, with propulsion mechanisms being due to locally generated surface processes \cite{anderson1989colloid, ebbens2010pursuit} or externally applied fields \cite{anderson1989colloid, snezhko2011magnetic, biswas2021rigidity}. The dynamics of extended bodies of these active colloids is a wide area of current research \cite{snezhko2011magnetic, vutukuri2017rational, zhang2016natural, nishiguchi2018flagellar, yang2020reconfigurable, biswas2021rigidity}, with various self-organized dynamical topologies reported - including linear \cite{vutukuri2017rational, reyes2023magnetic}, C-shaped \cite{vutukuri2017rational}, helical \cite{yan2012linking, vutukuri2017rational, biswas2021rigidity}, rings \cite{nishiguchi2018flagellar}, and rotating chiral clusters \cite{zhang2016natural}. 

% The fluid flow generated by each active particle mediates the interactions between these colloids and control their collective dynamics \cite{thutupalli2018flow, bechinger2016}, 
% while the motility of individual particles alone has been shown to result in phase separated states with distinct nonequilibirum characteristics \cite{cates2015motility}. 

Another parallel long-standing goal of soft matter science is the self-assembly rules of colloidal particles into targeted self-shaping structures \cite{bechinger2016, zeravcic2014size}. In these works, the object of interest typically is the \textit{structural} details (for instance the configuration) of the self-assembled colloidal molecule \cite{zeravcic2017colloquium}. In out-of-equilibrium systems, not only the emergent dynamics itself but also the ability to design a prescribed dynamical function into the collective dynamics \cite{soto2014self, bishop2023active} is of interest.

A specific class of colloidal interactions are those that are mediated by a chemical trail \cite{sengupta2009dynamics, taktikos2012collective, hokmabad2022chemotactic}. These interactions have also been widely reported in various biological systems - including bacteria \cite{mittal2003motility} and ant colonies \cite{jackson2006longevity}, with demonstrated roles in biofilm \cite{gelimson2016multicellular} and colony \cite{zhao2013psl} formation. Theoretically, these have been studied at the individual \cite{kranz2016effective}, and collective \cite{taktikos2012collective, gelimson2016multicellular} levels. Recently, it has been demonstrated (experimentally) that the trail mediated chemo-repulsive interactions between the monomers of an active colloidal chain results in an emergent C-shape rigidity of the chain \cite{kumar2023emergent}.

\begin{figure} 
    \centering
    \includegraphics[width=.80\textwidth]{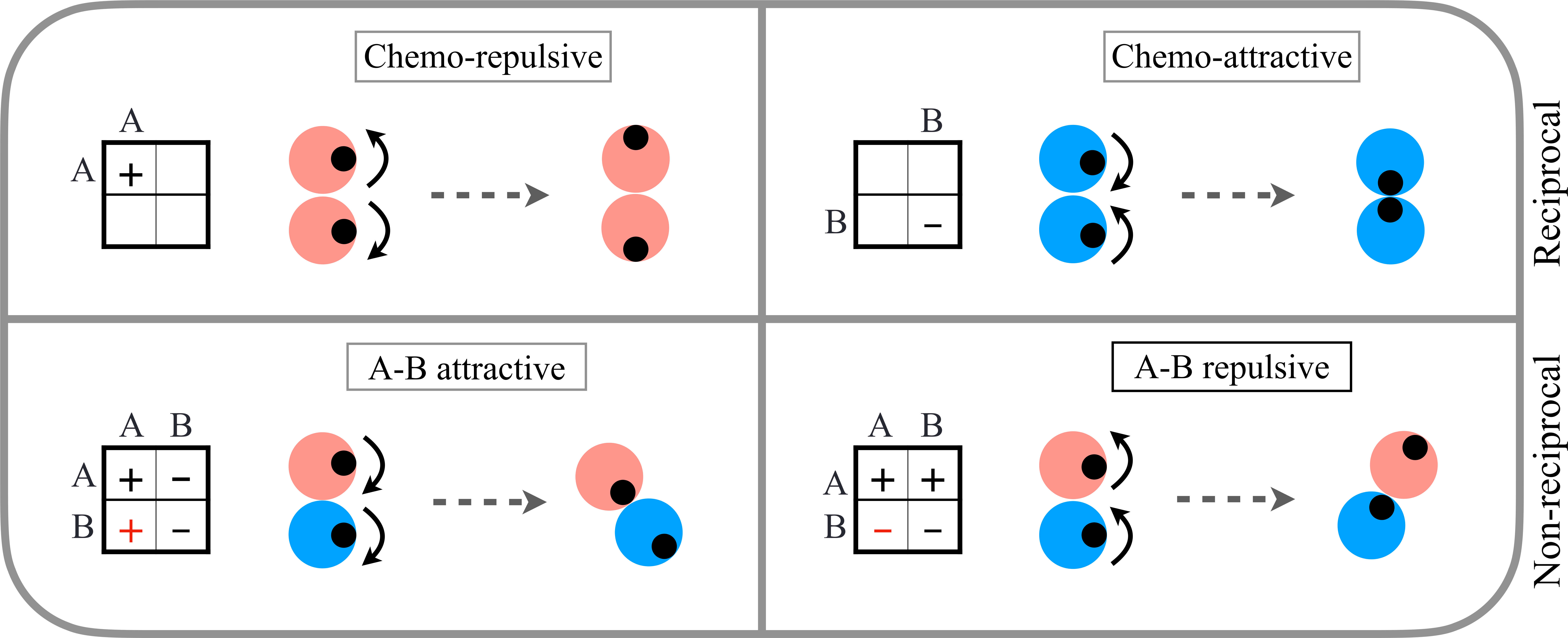}
    \caption{{Schematic diagram for chemical interactions of a dimer}.
    Two species of A type (denoted by red circles) are purely repulsive, while those of B type (blue circles) are purely attractive. The monomers are free to rotate and slide on each other, while the distance between the monomers does not change because they are freely-jointed. The black dot on the circles indicate the orientation of the particles, while curved arrows show rotations. For repulsive interactions, monomers preferentially point away from each other, whilst for attractive interactions they preferentially point towards each other. For a dimer of both A and B types, the interactions are non-reciprocal (\textit{bottom row}). There are two choices: A attracts B and B is repelled away from A and vice-versa. 
    {Chemical affinity matrices (refereed to as $\mathbf{\Upsilon}_{ij}$ in Eq.\eqref{eq:curr})} in the inset imply attraction (-) or repulsion (+) between monomers. These matrices specify the nature of both self-interaction as well as inter-particle interactions.  
    %Note that the angular velocity $\mathbf{\Omega}$ depend both on position and orientation.
    }
    \label{fig1}
\end{figure}

In this paper, we study the trail-mediated collective dynamics of a collection of chemically self-interacting active colloids joined by harmonic springs - which we term an \textit{active colloidal chain}. We cover the full range of chemo-repulsive, chemo-attractive, and non-reciprocal chemical interactions between the monomers of the chain. A summary of the interaction types studied here is shown in Fig.~(\ref{fig1}). Our model, consisting of two basic ingredients - trail-mediated chemotactic interactions between monomers and internal spring forces - we show to be able to support a wide range of phenomenology. In monodisperse chains, we report ``universal" C-shape flocks in the chemo-repulsive case, with the universality arising via the forward-backward symmetry breaking due to the presence of chemical trails, and crystallization in the chemo-attractive case. In a bidisperse chain (made of two species of monomers shown in Fig.~(\ref{fig1})), we show that undulatory gaits of purely chemical origin are supported, in addition to chiral molecules with tunable dynamics. Our model thus uncovers a role of trail-mediated interactions in sustaining a spatio-temporal asymmetry in colloidal dynamics, whilst providing a simple yet scalable model of tunable non-equilibrium self-assembly of active colloids, where the internal interactions have a non-reciprocity.

The rest of the paper is organized as follows. In Section \ref{sec:2}, we introduce our model of a chemically interacting chain of active colloidal particles, and specify important length and time scales, as well as dimensionless parameters that arise. In Sections \ref{sec:chemo-rep} and \ref{sec:chemo-attract} of the paper, we study the dynamics of the system via relevant length and time scales that govern the overall qualitative behaviour; along with time-averaged order parameters that provide quantitative-predictions. 
We obtain a full phase diagram of chemo-repulsive and attractive cases, with different regions delimited by different relative contributions of these length and time scales. 
Then, in Section \ref{sec:chemo-attract-rep}, we study a chain made of two species of monomers with non-reciprocal chemical affinities. In these chains, we report the existence of novel emergent dynamics - sustained oscillations, reversal of motion, and undulatory propulsion (a swimming gait). In Section \ref{sec:rotors}, we show how specifically chosen initial configurations and/or conditions can lead to a chirality in the dynamics. We conclude by suggesting how such assemblies can be used to achieve a programmable dynamics, where the structural and dynamical details of the colloidal molecule can be engineered by tuning the interaction details between species in the chain and/or the initial conditions. Finally, we discuss implications of our work and potential future studies.
% For active chains made of two and three monomers, we obtain analytical results for the centre of mass motion of the chain by reducing the the number of equations describing the dynamics using the symmetries of the system.
%%------------------------
%%------------------------

% The rest of the paper is organized as follows. In Section \ref{sec:2}, we introduce our model of the chemically interacting chain of active Brownian particles, and specify important length and time scales, as well as dimensionless parameters that arise. In Section \ref{sec:chemo-rep}, we discuss the chemo-repulsive case. We first outline our theoretical approach used in this problem, where we exploit the spatial symmetries of the chain to obtain the orientational fixed points and reduce the effective dimensionality of the problem. Then, we present results for the chemo-repulsive case, where we quantify both positional and angular rigidity of the chain. In Section \ref{sec:chemo-attract}, we present results for the chemo-attractive case, where we report collapsed crystalline lattices that (effectively) come to a halt. In Section \ref{sec:chemo-attract-rep}, we present results for the dimer and trimer cases, where we extend the system to a bidisperse one, and thus, allowing for the interactions between the monomers to be non-reciprocal. The result of non-reciprocity gives rise to novel collective phenomenon such as undulatory motion and reversal of direction of motion. Finally, in Section \ref{sec:summary}, we show that beyond trimers, molecules that exhibit chirality and/or specific structural features are attainable from this system, hinting to the possibility of design principles for such systems. Finally, we conclude and discuss potential future work.
% In what follows, we detail our model and describe our results. 

\section{Model}\label{sec:2}
\subsection{Equations of motion}
We model the $i$th active particle as a colloid particle centered at $\mathbf r_i=(x_i,y_i)$, confined to move in two-dimensions, which self-propels with a speed $v_s$,  along the directions $\mathbf e_i =(\cos\theta_i,\,\sin\theta_i)$. Here $i=1,2,3,\dots,N$.
The direction of the $i$th particle, given by the angle $\theta_i$, changes due to coupling its dynamics to the dynamics of a phoretic (scalar) field $c$, as we describe below. The position and orientation of the $i$th particle is updated as:
\begin{align}
\frac{d \mathbf r_i}{dt} =  {\mathbf V}_i,\quad\qquad \frac{d \mathbf e_i}{dt} = {\mathbf \Omega}_i \times \mathbf e_i. 
\label{eq:dyn}
\end{align}
Here, the translational velocity $\mathbf V_i$ and angular velocity $\mathbf \Omega_i$ of the $i$th particle are given as:
% \begin{subequations}
 \begin{align}
\label{eq:dyn2}
          {\mathbf V}_i &= v_s \mathbf e_i +\chi_t\, \mathbf J_i + \mu \mathbf F_i 
   + \sqrt{2D_t}\,\bm\xi^T_i
    ,\qquad\quad
    \mathbf \Omega_i %&
    = \chi_r 
    \left(\mathbf e_i\times\mathbf {\mathbf{{J}}}_i 
    \right) 
   + \sqrt{2D_r}\,\bm\xi^R_i.
 \end{align}
% \end{subequations}
In the above, $\mu$ is mobility, $D_t$ and $D_r$, are respectively, translational and rotational diffusion constants of the particle, while $\bm\xi_i^T$ and $\bm \xi_i^R$ 
are white noises. 
The chemical interaction between the particles is given in terms of the chemical current $\mathbf {\mathbf{\mathbf{{J}}}}_i$, which is described in detail below.
The chemical interactions in our model (\ref{eq:dyn2}) drive  orientational changes through the terms proportional to 
$\chi_r$, and positional changes through the term proportional to $\chi_t$.
% The `translational chemotactic' term (e.g. present in \cite{hokmabad2022chemotactic, saha2014clusters, sengupta2009dynamics}) is not present in our model, as it does not alter the dynamics due to the presence of spring connections between the monomers of the chain. 

The force keeping the chain together is modelled as a spring. The body force on the $i$th particle is given as: $\mathbf F_i = -\mathbf\nabla_i  \mathcal U$, while the potential $\mathcal U$ is:
$\mathcal U=\sum_{i=1}^{N-1} \mathcal U^s (\mathbf r_i,\mathbf r_{i+1})+\sum_{i<j}\mathcal U^e   (\mathbf r_i,\mathbf r_{j}).$
Here, $\mathcal U^s=k_{sp}\left(r_{ij}-2b\right)^2$ is spring potential of stiffness $k_{sp}$ and natural length $2b$ which holds the chain together and only acts between neighbouring particles in the chain. $r_{ij}=|\mathbf r_i - \mathbf r_j|$ and $b$ is the radius of the monomers.
$\mathcal U^e$ is a repulsive potential precluding overlap of particles. We choose it to be also of the form $\mathcal U^e=k^e\left(r_{ij}-2b\right)^2$ if $r_{ij}<2b$, while it vanishes otherwise. $k^e$ is a constant, which determines the strength of the repulsive force, which precludes overlap of particles. Note that there is no bending potential added. Thus, we expect the chain to be highly flexible. 
We show below (Section \ref{sec:chemo-rep}) that chemical interaction between the monomers of the chain leads to an emergent rigidity in the chain. 

The chemical interaction between the particles is given in terms of chemical current $\mathbf {\mathbf{\mathbf{{J}}}}_i$ (see Eq.~(\ref{eq:dyn2})), which is defined as: 
$\mathbf {\mathbf{\mathbf{{J}}}}_i(t) =-\left[\mathbf\nabla c(\mathbf r, t)\right]_{\mathbf r=\mathbf r_i}$,  
where $c(\mathbf r, t)$ is the concentration of filled micelles at the location $\mathbf r$ at time $t$. In our model, where each particle is considered as a point source of chemicals, the dynamics of the field $c$ follows from the equation:
 \begin{align}
    \frac{\partial c (\mathbf r, t)}{\partial t} %+ \mathbf v\cdot \nabla c
    = D_{c}\nabla^2 c(\mathbf r, t) + \sum_{i=1}^Nc_0\,
    \delta(\mathbf r- \mathbf r_i(t)).
    \label{le2}
\end{align}
Here $D_{c}$ is the diffusion coefficient of the filled micelles, $c_0$ is emission constant of the micelles, and $N$ is the number of monomers in the chain. It should be noted that we model the particle as a point emitter of chemicals. Using the above, we can solve for $\mathbf{\mathbf{\mathbf{{J}}}}_i(\mathbf r_i, t)$ as:
\begin{align}
    \mathbf J_i(\mathbf r_i,t) = \frac {c_0}{\pi b}
    \sum_{\substack{j=1
    % \\i = j
    }}^N 
    \left[\bm\Upsilon_{ij}%\odot
    \int\limits^{t-t_0}_{0} dt'
    \left(
    \frac{\Delta\mathbf r_{ij}  }{\mathcal D^2}
    \right)
    % {\mathcal{F}_{ij}(t,t')}
    \exp \left(
    -\frac{\left[\Delta\mathbf r_{ij}\right]^2  }{\mathcal D}
    \right)
    \right],\qquad  \Delta\mathbf r_{ij}  = \mathbf {r}_i(t) - \mathbf r_j(t').
    \label{eq:curr}
\end{align}
Here $\mathcal D=4D_c|t-t'|$, while 
$t_0$ is a finite time delay between the chemical being released and sensed, which preempts any divergence in the mathematical form of $\mathbf{J}_i(t)$ \cite{sengupta2009dynamics}. In simulations, we choose $t_0$ to be much smaller than any other time-scale in the problem.
The tensor $\mathbf\Upsilon_{ij}$, or the  chemical affinity matrix, contains our choice of chemical affinities between different particles $i$ and $j$ (both  self-interaction as well as inter-particle interactions).
%while $\odot$ implies maximal contraction between two tensors.
{The case of $\Upsilon_{ij}<0$ (the \textit{chemo-attractive} case) corresponding to monomers rotating/translating towards each other, while $\Upsilon_{ij}>0$ (the \textit{chemo-repulsive} case) implies that the particles rotate/translate away from each other. This scenario corresponds to the well known mechanism of repulsion due to the trails of other monomers in the system \cite{herminghaus2014interfacial, hokmabad2022chemotactic, kumar2023emergent}. }
The  chemical affinity matrix ($\Upsilon_{ij}$) can also suitably be chosen to design non-reciprocal affinities, as we explain in Section (\ref{sec:chemo-attract-rep}) below.
A pictorial representation of $\mathbf\Upsilon$ for dimers is in the inset matrices of Fig. \ref{fig1}. 
{The above model (without the  chemical affinity matrix $\Upsilon_{ij}$) was used in a recent work \cite{kumar2023emergent} to capture experimental phenomenology with monomers having chemo-repulsive interactions. 
In this work, we do an extensive study of chemo-repulsive interactions and map the phase diagram and obtain newer theoretical results. We also extend the study to a chain of chemo-attractive particles. In the model, we have now introduced the interactions matrix which also allow for monomers of different kinds in the same chain. Thus, in addition, we study non-reciprocal chemical self-interactions (see Figure \ref{fig1}).}

\subsection{Length and time scales}
We will now define some length scales and time scales that arise from this model. The basic length scale here is the monomer radius, $b$. This amounts to an additional length scale, imposed via excluded volume repulsion, that appears in Eq. (\ref{eq:dyn2}) via $\mathbf{F}$. This is contrasted with ``natural" length scales, that arise from non-dimensionalizing the equations of motion. There are three natural length scales:
\begin{align}
l_c = \frac{D_{c}}{v_{s}}, \quad l_r = \frac{\chi_r}{b^{2} D_{r}}, \quad l_t = \frac{\chi_{t}}{D_{t}}.
\label{eq:lscales}
\end{align}
$l_c$ gives the characteristic distance at which the chemical (i.e filled micelles) diffuses across the system, whilst $l_r$ gives the characteristic amount by which monomers change their orientation in response to the chemical gradients (as opposed to spontaneously, which is simply $b$). $l_{t}$ is the typical distance colloids translate deterministically in response to chemical gradients.
There are four time scales \footnote{There are additional time scales $\tau_{\scriptstyle{d}}=\frac{D_{c}}{v_{s}^{2} }$ and 
$ \tau_t = \frac{b^{2}}{\chi_t}$. We
ignore these in the analysis.}, they are:
\begin{align}
    \tau = \frac{b}{v_{s}}, \,\,\,\,\,\,
    \tau_{r} = \frac{1}{D_{r}}, \,\,\,\,\,\,
    \tau_f = \frac{b^{3}}{\chi_r}, \,\,\,\,\,\,
    \tau_{c} = \frac{b^{2}}{D_{c}}.%, \,\,\,\,\,\,
\label{eq:tscales}
\end{align} 
 Here, $\tau$ is a spontaneous propulsion time scale, which is the time it takes for an isolated particle to move a distance equalling its radius in absence of any reorientation, $\tau_{r}$ is the time at which orientations turn randomly due to noise, $\tau_f$ is a deterministic time during which the orientations change in response to the chemical field, $\tau_{c}$ is the time scale at which chemicals diffuse across the system.
 %, and $\tau_t$ is the time scale by which colloids deterministically propel a distance equalling their radius from chemical interactions between them.

\subsection{Dimensionless parameters}
Let us define two dimensionless activity parameters which we will use for the study:
\begin{align}
    \mathcal{A}_{1} &= \frac{bv_{s}}{ D_{c}}=\frac{b}{l_{d}} = \frac{\tau_{c}}{\tau}, \qquad
    \mathcal{A}_{2} = \frac{\chi_r}{b^{3} D_r}
    = \frac{l_r}{b} = \frac{\tau_{r}}{\tau_f}.
\label{eq:peclet_nums}
\end{align}
Here we have used equations (\ref{eq:lscales}) and (\ref{eq:tscales}).
%as $\mathcal{A}_{1} = $, and $\mathcal{A}_{2} =. 
$\mathcal{A}_{1}$ captures the relative effect of spontaneous propulsion versus diffusion of filled micelles in the system, which in turn induces the \textit{deterministic} rotation. Intuitively, for a sufficiently large $\mathcal{A}_{1}$ the monomers evade the diffusing chemicals and no chemically-induced self-orgnanization will be possible, whereas below a threshold, the chemicals are sufficient diffused such that the chemical contribution is evident. We will later see in Sections \ref{sec:chemo-rep} and \ref{sec:chemo-attract} that this intuition is reproduced in the simulations. 
 $\mathcal{A}_{2}$ on the other hand captures the relative contribution of deterministic versus random rotations of the orientations. The relative contributions of these various length and time scales, where appropriate, will be discussed in the context of the qualitative behaviour of the chain, in Sections \ref{sec:chemo-rep} when discussing the phase diagram 
 \footnote{With the definitions given in equations (\ref{eq:lscales}) and (\ref{eq:tscales}) there are additional valid (10 in total) Peclet numbers. However, these additional feasible dimensionless parameters in our system (e.g. one could define $\mathcal{A}=\frac{v_{s}b^{2}}{\chi_r} = \frac{\tau}{\tau_f}$), are not systematically studied here - indeed they do not affect our bigger picture findings}.

%%------------------------------------------------
% \section{Results and discussions}
\section{An active polymer of chemo-repulsive monomers}\label{sec:chemo-rep}
%%------------------------

We first study a chain made of only one kind of monomers, which are chemo-repulsive in nature. The $\Upsilon$ matrix here has the form $\Upsilon_{ij} = 1 \quad \forall i,j$. We study the phase behaviour of the system as a function of $\mathcal{A}_{1}$ and $\mathcal{A}_{2}$.  We note that, throughout the paper, we vary $\mathcal{A}_{1}$ by varying $D_{c}$ at a fixed $v_{s}$, whilst we vary $\mathcal{A}_{2}$ by varying $D_{r}$ at a fixed $\chi_r$.  Our results for a chain made of chemo-repulsive monomers are summarized in the phase diagram of Fig.~(\ref{fig_cr}), {delimited  by the curvature of the steady state chain (see appendix \ref{app:curv}).}
%%
%%------------------------
\begin{figure} 
    \centering
    \includegraphics[width=.94\textwidth]{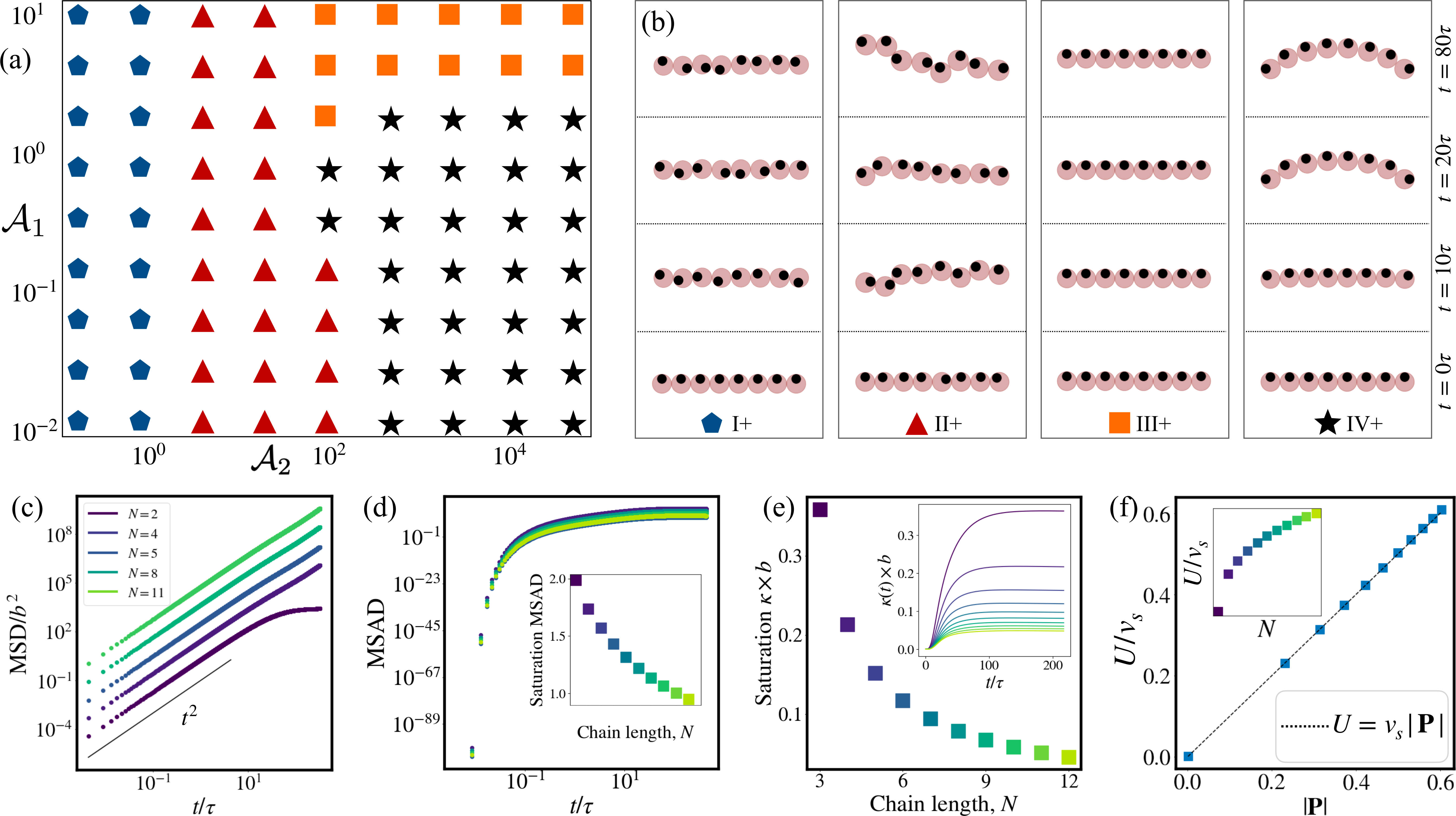}
    \caption{{Dynamics of a chain with chemo-repulsive monomers}.
    (a): the phase diagram (for $N=8$) in the $\mathcal A_1$-
    $\mathcal A_2$ plane. Panels in (b) show representative configuration from each distinct phase: (i) I+: Stiff and stationary chain, (ii) II+: Floppy chain (FJC-like dynamics), (iii) III+: Stiff and propelling state, (iv) rigid chain, self-propelling with C-shape.
    %Phase diagram is constructed for $N=8$. 
    Remaining panels are results for the phase IV+. Note that same color scheme is used across panels (c-f) and their insets to indicate number of monomers in the chain.  
(c): the ballistic nature of the polymers are captured via the MSD ($N=2$, $4$, $5$, $8$, and $11$ shown), for all except the dimer where the motion stops (see section \ref{sec:dimerMono}). 
Note that the lines are vertically adjusted such that there is spacing in between for distinguishability.
    The saturation value of curvature ($\kappa$) of the chains as a function of $N$ is shown in (d), with the time evolution of $\kappa$ shown in the inset. 
    %It can be seen that longer chains are stiffer. 
    In (e), the MSAD of the chains are shown. In the inset the saturation value is displayed, indicating now an ``angular stiffening" of the chain with $N$. In panel (f) relation (\ref{eq:u_vs_pol}) for the speed ($U$) of the chain is verified, whilst its value with $N$ is shown in the inset. 
}
    \label{fig_cr}
\end{figure}
%%------------------------
\subsection{Phase diagram} Region I+ corresponds to a stiff-noisy chain (see Fig.~(\ref{fig_cr})) blue symbol, also bottom panel \textit{top row}). This is defined by a rapid $\tau_{r}$ with respect to $\tau_f$, where $\frac{\tau_{r}}{\tau_f} = \mathcal{A}_{2} \ll 1$. In this region, we also have that $\frac{l_r}{b} = \mathcal{A}_{2} \ll 1$ (thus distances where the chain could potentially fold are negligible compared to the monomer radius). 
Note also that in this region $\tau_r / \tau_{f} \sim \mathcal O(10^{-3} - 10^{-1}) \ll \tau / \tau_{f} \sim \mathcal O(1)$, thus the chain orientations fluctuate much quicker than they can pick up a direction of motion. 
%It is worthwhile to note that the monomers can orient in any direction randomly, and indeed point backwards. See SI Video 1 \cite {siText}.
Increasing $\mathcal{A}_{2}$, we arrive at Region II+, which corresponds to a quasi-FJC (freely-jointed chain) region, shown in Fig.~(\ref{fig_cr}), red symbol. 
Here, the chain translationally diffuses, but folds into and out of a given configuration (see Fig.~(\ref{fig_cr}) bottom panel \textit{second to top} row). 
In this region we have that $ \mathcal{A}_{2}  \sim \mathcal O(10^0 - 10^2)$, i.e. the diffusive ($\tau_r$) and deterministic ($\tau_f$) impart equal contributions of deterministic and diffusive rotational motion - resulting in the motion qualitatively resembling a freely-jointed chain  - in that it neither translates nor folds deterministically \cite{doi1988theory}. Region III+ is denoted in Fig.~(\ref{fig_cr}) with the yellow squares. This is the weakly interacting region, where the chain propels deterministically, but does not fold (see Fig.~(\ref{fig_cr}), bottom panel \textit{second to bottom} row). Here, we have that $\frac{\tau_{c}}{\tau} = \mathcal{A}_{1}^{-1} \ll 1$, thus the chemical diffuses much slower that the spontaneous propulsion time,  and equally $\frac{l_{d}}{b} = \mathcal{A}_{1}^{-1} \ll 1$. Finally, region IV+, denoted with black symbols in Fig.~(\ref{fig_cr}), is the C-shape region, where the polymer propels deterministically and self-organizes into a stable configuration that resembles the alphabet C (thus acquiring a finite curvature), with a continuously varying orientation across the chain and an overall rigidity \cite{kumar2023emergent, vutukuri2017rational} (see Fig.~(\ref{fig_cr}), bottom panel \textit{bottom} row) . We have that $\frac{\tau_{r}}{\tau_f} \gg 1$ (diffusive rotational time is negligible), $\tau_{r} \sim \tau$ (both deterministic rotation and propulsion occur simultaneously), and $\frac{l_{d}}{b} \gg 1$.

The characteristic dynamics of each region is shown in Fig.~(\ref{fig_cr}) (\textit{bottom panel}) (Videos are in SI Video 1 \cite {siText}). After $t = 10 \tau$, we see that the orientational changes in the chain have set in. The dominant time scales here are $\tau_{r}$ for noise-dominated phases I+ and II+, whereas for phase IV+, the time-scale $\tau_f$ is dominant, with deterministic alignments seen. After $t \sim 20\tau$, we see that in phase IV+ steady-state spatial structures have formed, namely an acquisition of curvature to form a C-shape, which is identical at $t=80 \tau$. 
This C-shape is indeed acquired for any initial conditions (see SI Video 3 \cite{siText}), and thus the solution is said to be \textit{universal}. This result is a consequence of the \textit{history dependence} (a.k.a trail-mediation) of the dynamics (c.f. Eq.~(\ref{eq:curr})), in that the colloids preferentially point away from each other both due to their instantaneous interactions at time $t$ as well as the collective memory of the past trajectories $t' < t$.
This historical dependence washes any symmetry from the initial condition of the monomer positions and orientation in the chain. In absence of historical chemical interactions, the chain will preserve the initial symmetry in the steady-state dynamics. 
{To rationalize the C-shape, note that the central monomer(s) always align perpendicular to the body axis (along the propulsion direction). The monomers away from this are repelled in proportion their distance from the center. Thus, the only possible (qualitative) shape that is selected is the C-shape, with a leading center and trailing edges.}
% {The C-shape is exclusively selected via the following mechanism. The orientation of monomers near the center of the chain will always be aligned perpendicular to the body axis, and hence along the propulsion direction. This is due to their mutual repulsion from monomers either side.
% Moving away from the central monomer, orientations are further repelled from this central direction, in proportion to their distance from the central monomer.
% Thus, the spatial distribution of the chain is such that the central parts of the chain lead the edges, with continuously varying monomer orientations, giving the C-shape.}
% Naturally, this requires $\tau_{c}$ to be much smaller than other deterministic time scales $\tau$ and $\tau_f$, which is the case in phase IV+. In the absence of such a dominant $\tau_{c}$ - there is no historical buildup of the chemicals - we get phase III+, where no such universality is seen. 
This history-dependence of the interactions between the particles distinguishes this model (see also \cite{sengupta2009dynamics, hokmabad2022chemotactic}), from other chemically interacting colloidal models \cite{tsori2004self, soto2014self, agudo2019active, saha2014clusters}, where only instantaneous chemical interactions are considered.

%A study of the structure in phase IV+ is given in appendix \ref{app:phase4+}.

%%%-----------------------------
\subsection{Dynamics in Phase IV+}\label{app:phase4+}
%%%-----------------------------
We now study the phase IV+, specifically the noiseless case. In the stable C-shaped phase, dynamics can be further characterized by MSD (mean-squared displacement), which is given as MSD$(t) = \left\langle[ \mathbf{r}_i (t) - \mathbf{r}_i (0)]^{2} \right\rangle$ and MSAD (mean-squared angular displacements), which is defined as:
MSAD $(t) = \left\langle [ \theta_{i}(t) - \theta_{i}(0)]^{2} \right\rangle.$
% % \begin{subequations}
% \begin{align}
%     \text{} ,\qquad
%     \text{MSAD} 
% \label{eq:msd_msad}
% \end{align}
% % \end{subequations}
Here $\langle\dots\rangle$ denotes averaging over the monomers. 
We see in Fig.~(\ref{fig_cr}c) that the chain propels ballistically, $\text{MSD} \propto t^{2}$ for all values of $N \geq 3$. We note that for $N=2$ the chain stops, this will be explained in the Section \ref{sec:chemo-attract-rep} below. The positional stiffness of the chain is quantified by curvature $\kappa$ in Fig.~\ref{fig_cr}(d). 
We see that the initially stiff chain acquires a curvature throughout the dynamics, saturating at a particular value after $\sim 10^{2}s$. $\kappa$ saturates as a function of time to a $N$-dependent value in the steady-state. We plot saturation value of $\kappa$ as a function of $N$.  We see that for longer chains $\kappa$ decreases and eventually saturates to a finite value, indicating ``stiffer" longer chains. 
% In panel (b), 
Further, the angular stiffness is quantified by the $\text{MSAD}$ in Fig.~\ref{fig_cr}(e), which also saturates. The saturation MSAD (\textit{inset}) shows a similar decrease w.r.t $N$, indicating that longer chains are more ``angularly stiff" in that they deviate less from $\theta_{i}(0)$ throughout the dynamics. 

Next, we study the speed of the chain in the stable C-shape configuration. The speed of the chain can be studied by considering motion in the center of the mass (CM) frame of the references. We define the centre of the mass coordinate $\mathbf R$ and the \textit{polarization} vector $\mathbf{P}$ as:
\begin{align}
\mathbf{R}  = \frac{1}{N} \sum_{i=1}^N \mathbf{r}_{i},\qquad
        \mathbf{P} =\frac1N \sum_{i=1}^N  \mathbf{e}_i.
\label{eq:pol}
\end{align}
 The order parameter $\mathbf{P}$ is net \emph{polarisation} of the orientation of the monomers in the chain.
 It is useful to decompose the dynamics into mutually perpendicular components. Thus, we can define velocity components parallel to the chain alignment as:
% \begin{align}
$    \dot{r}_{i}^{\parallel} = \dot{\mathbf{r}}_{i}\cdot \mathbf{P} 
$.
% \qquad \mathbf{\dot{R}}_{i}^{\perp} = \dot{\mathbf{R}}_{i} \cdot (\mathbf{I} - \mathbf{p}\mathbf{p}^{T})
% \label{eq:vel_decomp}
% \end{align}
% With its center-of-mass 
% \begin{align}
% %, \qquad \dot{\mathbf{R}}^{\text{cm}, \perp} = \frac{1}{N} \sum_{i} \dot{\mathbf{R}}_{i}^{\perp}
% \label{eq:com_comp}
% \end{align}
We also define the average speed of the chain in this case as:
    $U = \langle\dot{R}^{\parallel}\rangle_{\text{ss}}$,
    and 
    $\dot{R}^{\parallel} = \frac{1}{N} \sum_{i=1} \dot{r}_{i}^{\parallel}$.  
Here $\langle\dots\rangle_{\text{ss}}$ implies that the average is taken in the steady-state. 
It is possible to reduce the 3N degrees of freedom given by equations (\ref{eq:dyn}), (\ref{eq:dyn2}) to re-cast the problem in terms of the time average; this dimensionality reduction procedure is explained in Appendix \ref{sec:SymmSec}. The result of this procedure yields the following equation:
\begin{align}
    U = v_{s} |\mathbf{P}|.
\label{eq:u_vs_pol}
\end{align}
The above is a linear relationship between  \textit{time-averaged quantity} $U$ and an order parameter $\mathbf{P}$ of this $3N$ dimensional system. Thus, although a linear relation akin to (\ref{eq:u_vs_pol}) is applicable at all times (using the instantaneous velocity instead of the time-average), the above is a unique specific case, relating a time-averaged dynamical quantity to an order parameter of the system. This is possible due to the symmetry of the dynamics as detailed in Appendix \ref{sec:SymmSec}. A plot of $U/v_{s}$ versus $|\mathbf{P}|$ is given in Fig.~\ref{fig_cr}(f). Here, the simulation data is predicted by Eq.~(\ref{eq:u_vs_pol}), validating our dimensionality reduction procedure (Appendix \ref{sec:SymmSec}). In  the inset, we see that the $U/v_{s}$ increases with $N$, corresponding to the behaviour of $\mathbf{P}$ in this limit.  Thus, longer chains, which are stiffer, also converge to a higher steady-state propulsion speed.

\section{An active polymer of chemo-attractive monomers}
\label{sec:chemo-attract}
%%-----------------------------------------------------
%%-----------------------------------------------------
We now turn to the case of chemo-attractive monomers. Each element of the interaction  matrix here is: $\Upsilon_{ij} = -1 \quad \forall i,j$. Here, the monomers will preferentially point \textit{towards} each other upon sensing the chemical field. We report a colloidal crystallization mechanism \cite{palacci2013living, cates2015motility, singh2016universal,thutupalli2018flow, taktikos2012collective} where the long-ranged attraction renders the initially propelling chain to undergo a collapse onto itself thereafter remaining quasi-stationary (see SI Videos 3, 5 \cite{siText}).

% \subsubsection{Metastable crystallites for $N<8$}

\begin{figure} 
    \centering
    \includegraphics[width=.884\textwidth]{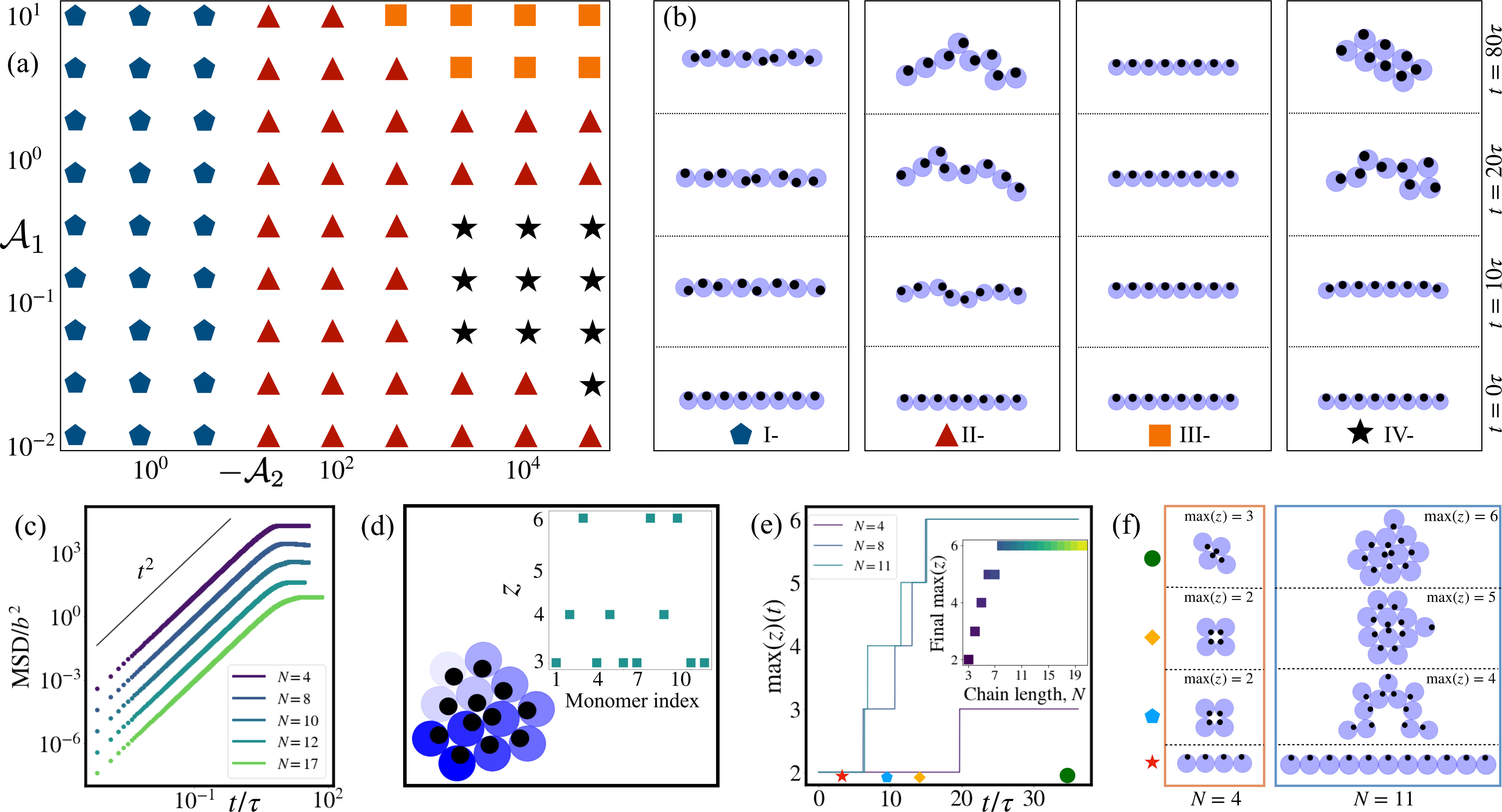}
    \caption{{{Dynamics of a chain with chemo-attractive monomers.}} Panel (a) shows the  phase diagram for $N=8$. The four phases are: (i) I-: Stiff + stationary chain, (ii) II-: FJC-like dynamics, (iii) III-: Stiff+propelling state, (iv) IV-: closed packed lattice. 
    %Phase diagram is constructed for $N=8$.
    Panel (c-f) show results for phase IV-.
    In (c), the MSD in plotted (for $N=4,8,10,12, \text{and} 17$), where the halting of motion is seen at $t \sim 10\tau$.
    (d) displays an example of a hexagonally-packed crystal. The colors are shaded such according to monomer number, with the vertically most upward being the lightest. The coordination number for the configuration is shown in the inset, with the central three ($z=6$), edge ($z=4$) and vertex ($z=3$) monomers displayed.  In (e), examples of the time evolution of our packing order parameter $\max(z)(t)$ is shown for $N=4$,$8$, and $11$, to distinguish between metastable the lattice of $N=4$ and the remaining two hexagonal packing. The former has a long lived transient lasting $O(10^{1}) \tau$, whilst the latter has much short-lived transients, eventually reaching $\max(z) = 6$. In the inset, the final value is shown, showing that $\max(z)=6$ is attained from $N=8$ onwards. For the time points labelled in the bottom of (e), the snapshots of the packing are displayed in (f), for $N=4$ (left) and $N=11$ (right).}
    \label{fig_ca}
\end{figure}
%
%
%%%===========================
%%%===========================
\subsection{Phase diagram}
%%%===========================
%%%===========================
{
For the chemo-attractive system, we use $\max(z)$ - maximum coordination number - to distinguish the different phases (see appendix \ref{app:coordZ}).
The resulting phase diagram is shown in Fig.~(\ref{fig_ca}).} 
It is qualitatively similar to that in Fig.~(\ref{fig_cr}), and thus, we have identically labelled the phases as I-, II-,III-, and IV-, where the aforementioned relative dominance of relevant length and time scales in each of the phases are identical. 
The region IV- differs from IV+ in that it now describe a collapsed lattice structure. Further, the line delimiting phases IV and II for large $\mathcal{A}_{1}$ (the \textit{weakly interacting} region) if shifted, such that large $\mathcal{A}_{1}$ and large $\mathcal{A}_{2}$ (e.g. $(\mathcal{A}_{1},\mathcal{A}_{2}) = (10^{3},10^{-2})$) does not crystallize. 
We expect that a similar trend to be seen for the chemo-repulsive case if values $\mathcal{A}_{1} \ll 10^{-2}$ are probed - i.e no stable C-configuration when the system is weakly interacting since there are no discernible concentration gradients in the system. 
Phases II+ and II- are also qualitatively different, in that in the latter case, one sees more distinct ``FJC-like" behaviour, with the chain collapsing and relapsing on itself, compared to II+ where there is no collapse as such. Movies of the chemo-attractive phases are in SI Video 4 \cite{siText}. Rotational diffusion has previously been argued to drive crystal break-up and re-arrangement \cite{palacci2013living} in activity-driven aggregation \cite{theurkauff2012dynamic,   cates2015motility}, and we see a chain analogue of this in this phase. 
%In addition, we have also measured in the MSAD (not shown in this paper), where, unlike in the chemo-repulsive case, there is no stiffening with larger $N$, by reason of its collapsed configuration. 
%%
%%%
\subsection{Dynamics in phase IV-}
We first distinguish two separate cases of collapsed chains, which we call (i) \textit{metastable crystallites}, and (ii) \textit{hexagonally packed} crystals. The former is characterized by long-lived (typically $\sim 10^{1}-10^{2} \tau$)  transient metastable crystals (either propelling or stationary depending on net $\mathbf{P}$) that collapse onto other crystal within the time scale of simulation, whilst the latter does not display any such transient states (any transients here collapse within a time less than O($\tau$)). For all chain lengths $N$, the steady-state crystals are close-packed structures, such that each monomer has $z \geq 2$. For chains of $N \geq 8$, this is identified by $\max(z) = 6$ (hexagonal packing). For $N<8$, we in turn have the metastable crystallites. An example of this is shown in Fig. \ref{fig_ca}(f), with the left panel displaying the packing evolution of the $N=4$ chain, whilst the right panel displaying $N=11$. For the former, we see that the metastable ``square" structure, with $\max(z)=3$, exists at times $t = 10 \tau$ and $t = 13 \tau$ (top panel \textit{second} and \textit{third} columns, eventually closing into a ``cone" structure, with $\max(z)=3$, at $t=35 \tau$ (\textit{right} column). This metastable crystal is captured via the time evolution of $\max(z(t))$ in Fig.~\ref{fig_ca}(e), with only a discrete change in $\max(z(t))$ seen, corresponding to the ``square-to-cone" transition. The behaviour for hexagonally packed crystals are markedly different, as is also seen in Figs. \ref{fig_ca}(f) ($N=11$) and \ref{fig_ca}(e) ($N=8$ and $N=11$). In particular, we see that any transient lattices do not last longer than a few $\tau$, before packing into a more close-packed structure, eventually forming the crystal. 

Note that in theory $N=7$ permits hexagonal packing, though in our simulations this occurs for selected realizations of noisy simulations ($D_{r}>0$), or with carefully selected initial conditions that deviate from (\ref{eq:init_cond}). Thus, we do not study this hypothetical packed crystal any further. Videos of the collapse dynamics of the metastable crystallites are shown in SI Video 4 \cite{siText}. 
%These crystallites are studied in more detail in Appendix \ref{app:metastable}.

%%===========================
%%%===========================
% \subsubsection{Hexagonal packing}
%%%===========================
%%%===========================
An example of a hexagonally close-packed crystal is shown in Fig.~\ref{fig_ca}(d) for $N=12$. 
As a rule, either monomer on the edges of the initially straight configuration appear on the rim of the packed crystal. 
The specific set of $z$ for this $N=12$ configuration is shown in the inset, where the monomer set with $z=3$ correspond to those in the rim and part of a pair-vertex, the pairs being (in monomer index units) $(1,4)$, $(7,8)$, and $(11,12)$. 
Those with $z=4$ correspond to rim plus edge monomers, i.e neighbouring the vertex monomers on the same edge, whilst those with $z=6$ correspond to the inner monomers. Apart from that, the exact configuration of packing varies significantly with (among others) $\chi_{r}$. For instance, larger $\chi_{r}$ (thus larger $l_r$ that promotes folding stronger at the edges, since deterministic rotation dominates over longer length scales of the chain (versus propulsion) - see SI Video 5 \cite{siText}). We do not study any further the monomer distribution of the packed structure as a function of these parameters as part of this study. 

The dynamics of these structures display a quasi-halting of motion. This happens since the instantaneous $\dot{R} \propto |\mathbf{P} |$, and for the hexagonally packed structures $|\mathbf{P}|$ is vanishingly small for large $N$ (all monomers point towards each other), and so the final collapsed state will be quasi-stationary, with a marginal drift along $\mathbf{P}$. The halting of the chain is seen via the MSD in Fig.~\ref{fig_ca}(c), where a plateau is seen, which was absent in the chemo-repulsive case (with the exception of $N=2$). This occurs at $t \sim 10 \tau$, corresponding to the point of $\max(z)=6$ in Fig.~\ref{fig_ca}(e). The net polarization of this packed lattice is found to be of $\sim 10^{-2} $, with $U$ of the final crystal at most $1/10$th the minimum $U$ of the chemo-repulsive case. Movies of various packing sequences are shown in SI Video 5\cite{siText}.
In particular, the final packed crystals have no positional spacing in between next-to-nearest neighbouring monomers, which can be seen for $N=4$ of the first and second from bottom rows, and $N=11$ for the first from bottom row.

%%------------------------------------------------------------------------
\section{An active polymer of chemo-attractive and chemo-repulsive monomers}
\label{sec:chemo-attract-rep}
%%------------------------------------------------------------------------
We next turn our attention to a binary mixture system, where two species A and B populate the chain. Next, we focus on the case where the two-species interact in a non-reciprocal fashion, such that (for instance) A chases B, whilst B is repelled from A. 
Such non-reciprocal interactions are known to be present in (isolated) colloidal systems \cite{niu2018dynamics, yu2018chemical, schmidt2019light, grauer2020swarm, meredith2020predator, sharan2023pair}, and have been studied theoretically and computationally in various particle-based \cite{soto2014self, agudo2019active, saha2014clusters}; field-theoretic \cite{saha2020scalar, you2020nonreciprocity}, and continuum solid \cite{scheibner2020odd, braverman2021topological, tan2022odd} models. 
Our study here is distinct in that apart from non-reciprocity of chemical interactions \cite{agudo2019active, saha2020scalar, you2020nonreciprocity, scheibner2020odd}, 
our constituent units are themselves polar self-propelling particles as wells as they interact chemically with each other and their chemical trails owing to which we observe a wide range of phenomenology, many of which we believe are hitherto unreported in such systems.

In this Section, we will also study the dynamics of monodisperse dimers from the previous Sections \ref{sec:chemo-rep} and \ref{sec:chemo-attract} (i.e. the AA and BB configurations) - this will help in contrasting with results of the AB configuration. We first study \textit{dimers} of configurations AA,BB, and AB; and later proceed to the study of trimers. In the later case, we will show that the breaking of action-reaction symmetry in the monomer interactions lead to emergence of states that display various phenomena at the collective level: chirality, undulatory gaits and direction reversal. 
For the rest of this section, we will denote AB+(-) and that where species A is attracted to (repelled from) species B. For the rest of the paper, we study the noiseless case with sufficiently diffuse chemicals ($D_{r}=0$, with $\tau_{c}\ll\tau$, i.e in the Phase IV regime). 
%A full study of the different phases for dimers and trimers are not carried out here, though we expect the results to be qualitatively similar to those in the previous section. 

%%%_---------------------
\subsection{Dimers - single species}\label{sec:dimerMono}

\begin{figure}[t]
    \centering
    \includegraphics[width=.99\textwidth]{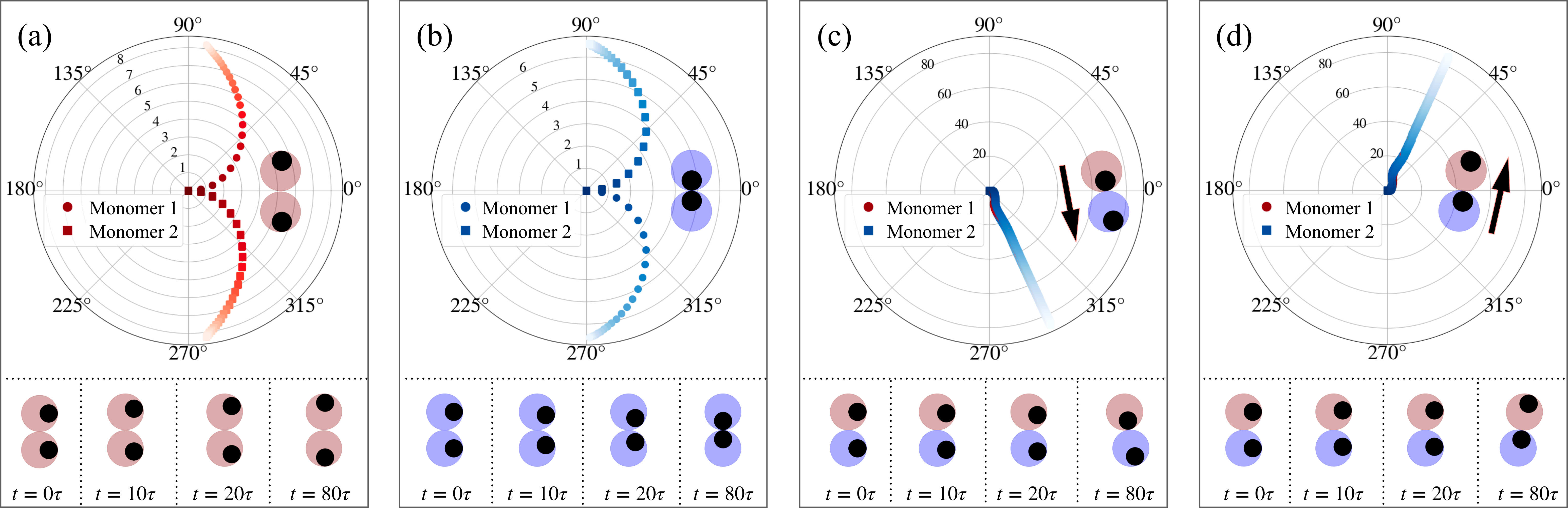}
    \caption{Dynamics of dimers for AA, BB and AB cases. Polar plots of the dynamics are displayed in panels (a) - (d). The radial axis of these plots indicate distance, $|\mathbf d_i(t) |/b = \sqrt{x_i(t)^{2} + y_i(t)^{2}}/b$, from the origin, angular axis indicates orientation for each monomer (using convention given in Appendix \ref{app:simDetailsParams}). The trajectories are color shaded according to time - from darkest (initial) to brightest (final). Inset indicates snapshot of steady-state dimer, with accompanying arrow showing its propulsion direction (no propulsion for AA and BB dimers in the steady-state).
     Each panels also displays snapshots at $t=0$, $10\tau$, $20\tau$, and $80 \tau$. 
}
    \label{fig6}
\end{figure}

 For the single species dimers, we have shown in Fig.~(\ref{fig_cr})  that the dimers (AA) come to a halt (see SI video 2 \cite{siText}). This is seen in the polar plot of Figure \ref{fig6}(a), where we see the initially vertical orientations reach the fixed points for $\theta_{1}^{*}=\frac{\pi}{2}$ and $\theta_{2}^{*}=-\frac{\pi}{2}$. This is also true for the BB configuration; this is shown in Fig.~\ref{fig6}(b), where now $\theta_{2}^{*}=\frac{\pi}{2}$ and $\theta_{1}^{*}=-\frac{\pi}{2}$. The orientational fixed points for the AA dimer are thus $\mathbf{e}_{1}^{*}=(0,1) $, and $\mathbf{e}_{2}^{*}=(0, -1)$.
For the BB configuration the fixed points are the opposite: $\mathbf{e}_{1}^{*}=(0,-1) $, and $\mathbf{e}_{2}^{*}=(0, 1)$
% It should be noted from Fig. (\ref{fig6}) that the AA and BB configurations are completely symmetric to one another.
We can use the above two equations to show that the dynamics of both AA and BB come to a halt. To this end, we have after adding Eq.~(\ref{eq:dyn}) for both species, 
\begin{align}
  \frac1N  \frac{d   }{dt} \left(\sum_{i}\mathbf{r}_{i}\right)=\frac{d   }{dt} \mathbf R=
  \frac1N v_{s} \sum_{i} \mathbf{e}^{*}_{i}   + \frac1N\mu \sum_{i} \mathbf F_{i}  =\frac1N v_{s} \sum_{i} \mathbf{e}^{*}_{i}. 
\end{align}
In the above we have used that fact that for $N=2$, $\mathbf{{J}}_{1} = - \mathbf{{J}}_{2}$ (rendered via the symmetries given in (\ref{eq:fpoints_even}), and for any $N$ the spring forces sum to zero). Thus,
$\sum_{i} \mathbf{e}^{*}_{i} = 0$, which implies that $\dot{\mathbf{R}} = 0$. Thus, we arrive at the result that the CM motion is static.

%%------------------------
%%%_---------------------
\subsection{Dimers - AB}
We now turn to the case of non-reciprocal dimers, where we uncover an ``alignment chasing" behaviour. Such two-species chasing has been reported in previous studies of colloidal models (far-field interaction with short ranged chemicals \cite{soto2014self}, field-theoretic models \cite{saha2020scalar, you2020nonreciprocity}, and prey-predator models \cite{tsyganov2003quasisoliton, sengupta2011chemotactic}. Thus, the alignment chasing reported here extends this phenomenon to an active colloidal chain in a straightforward manner. For the dimer-AB, the matrix entries for $\Upsilon_{ij}$ are given by:
% \begin{subequations}
\begin{align}
\Upsilon_{ij}^{AB} =\delta_{i1}\left(\delta_{j1}-\delta_{j2}\right)
- \delta_{i2}\left(\delta_{j2}  - 
%(1 - \delta_{i1}\delta_{j2}) 
  \delta_{j1}\right).
% \Upsilon_{ij}^{AB-} =  +\delta_{i1}\delta_{j2} - (1 - \delta_{i1}\delta_{j2}) \delta_{i2} \delta_{j1}. 
\end{align}    
% \end{subequations}
 We see that in the case of an AB-dimer, the action-reaction symmetry can be broken in two distinct ways [AB(+) and AB(-)]. We study both these cases below.

The results for the AA and BB configuration are shown in Figure \ref{fig6}(a) and (b) respectively. AB- configuration are shown in Figure \ref{fig6}(c), whilst the AB+ configuration is in \ref{fig6}(d). We see that in both AB+ and AB- cases, the monomers quickly settle on a solution where they are mutually aligned (having the same orientation) and self propel in a chasing behavior (see SI video 7 \cite{siText}). 

This chasing result is also straightforwardly deduced analytically. Considering the attractive AB interaction, we have that the fixed point for $\dot{\mathbf{e}}_{1_{A}} = 0 \implies 
\mathbf{e}_{1_{A}} \propto - |\chi_{r} | (\mathbf{r}_{1_{A}} - \mathbf{r}_{2_{B}})$.
On the other hand, $\dot{\mathbf{e}}_{2_{B}} = 0 \implies \mathbf{e}_{2_{B}}   \propto + |\chi_{r} | (\mathbf{r}_{2_{B}} - \mathbf{r}_{1_{A}})$. Thus, we note that $\mathbf{e}_{1_{A}}$ is parallel to $\mathbf{e}_{2_{B}}$. This happens after about $t \sim 80 \tau$ of the dynamics (see Fig.~\ref{fig6}(c), \textit{bottom row}.). On the other hand, the dynamics at the fixed point in the phase space of $\{ R_x, R_y \}$ read
\begin{align}
    \frac{dR_x}{dt} = 2v_{s}\cos(\theta^{*}),\qquad
    \frac{dR_y}{dt} = 2v_{s}\sin(\theta^{*}),
\label{eq:ab_red_fp}
\end{align}
after which the dimer propels indefinitely along a fixed angle $\theta^{*}$ which we can also deduce. The steady state angle the dimer picks is $\approx 292.5 ^{\circ}$, half way in between that of a full vertically downward chase ($\theta = 270^{\circ})$, and an unbiased diagonal $\theta = 315 ^{\circ}$, owing to the symmetry of the problem. For the AB(+) case, the same argument gives the steady-state angle as $\approx 67.5 ^{\circ}$. 

Let us summarize the results for dimer interactions. The monodisperse dimers (both AA and BB) come to a halt, whereas the bidisperse dimers propel at a fixed angle, one species chasing the other indefinitely. The reduced phase space is 3 dimensional for the monodisperse dimers \textit{throughout} the dynamics (see Appendix \ref{app:dim_red_mono} for details of the dimensionality reduction procedure). For the AB molecules, the reduced phase space is also 3 dimensional, whilst it is 1 dimensional \textit{at the fixed point orientations}, corresponding to a stably propelling dimer. With this, we now move on to the trimer cases.

%%------------------------
\begin{figure}
    \centering
    \includegraphics[width=.95\textwidth]{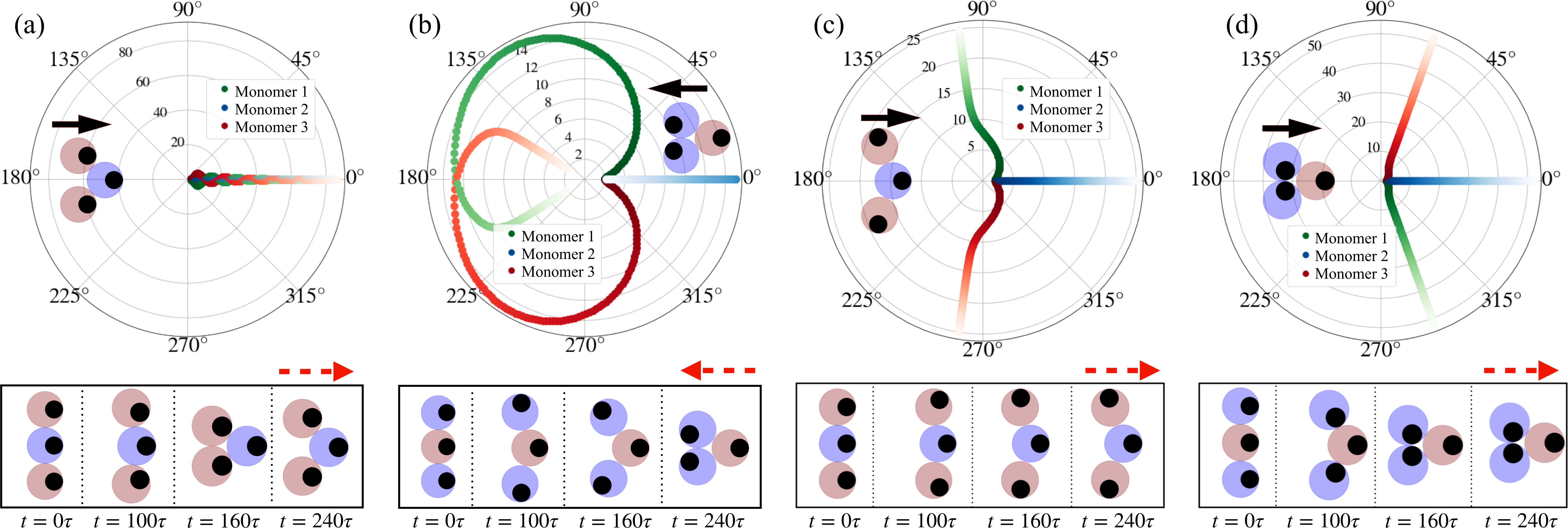}
    \caption{Dynamics of trimers  with chemical affinities that break the action-reaction symmetry. Panels (a) and (b) show those of the (-) configurations whereas (c) and (d) on the right are for the (+) configurations. Panels (a)-(b) on top row display polar plots for ABA(-), BAB(-) respectively. Black arrow indicates direction of $\dot{R}$. The bottom panels show the snapshots from simulations for the corresponding cases. Red dotted arrow on right most snapshot indicates net polarization direction. (c) and (d) are identical to (a) and (b) respectively but for the ABA(+) and BAB(+) configurations.
    }
    \label{fig7}
\end{figure}
\subsection{Trimers: oscillations, swimming, and reversal of motion due to breaking of action-reaction symmetry}
Having established that dimers perform deterministic chase, we now study the trimer chains, specifically the configurations of ABA, BAB, and BBA. We uncover striking consequences as a result of this additional degree(s) of freedom of the trimer. In particular, chasing as in the previous section is now extended to an undulatory gait for BBA(-) and AAB(+), and in addition we will obtain coupled oscillations of ``tail" monomers for ABA(-). Finally, we will see that BAB(-) \textit{reverses} its direction of motion. 
For these configurations, the matrix entries for $\Upsilon_{ij}$ are given by:
% \begin{subequations}
\begin{align}
    \Upsilon_{ij}^{ABA} =& 
   [\delta_{i1}+\delta_{i2}+\delta_{i3}][\delta_{j1} - \delta_{j2} + \delta_{j3}],
   %\\ &+ (\delta_{j1} -\delta_{j2} + \delta_{j3}) 
   \quad 
   % &+
   % \delta_{i3}(\delta_{j1} -\delta_{j2} + \delta_{j3}) \\
    \Upsilon_{ij}^{BAB} =  - \Upsilon_{ij}^{ABA},\qquad
    \Upsilon_{ij}^{BBA} = [ \delta_{i1}+\delta_{i2}+\delta_{i3}][ - \delta_{j1} - \delta_{j2} + \delta_{j3}].
\label{eq:trinary_matrix}
\end{align}    
% \end{subequations}

We further quantify the dynamics by the bond angle $\phi$ of the trimer. This uniquely specifies the relative monomer positions for the ABA and BAB configurations. We define an ``angular momentum" of the system
\begin{align}
    L_{z} = \frac{1}{N} \sum_{i}^{N} \left( \mathbf{V}_{i} \times ( \mathbf{r}_{i} -\mathbf{R}) \right) \cdot \mathbf{\hat{z}}
\label{eq:ang_mom}
\end{align}
which quantifies net rotations of the monomers about the center of mass (for a 2D system there is only one such component projecting out of plane of the page).

\begin{figure*}
    \centering
    \includegraphics[width=.90\textwidth]{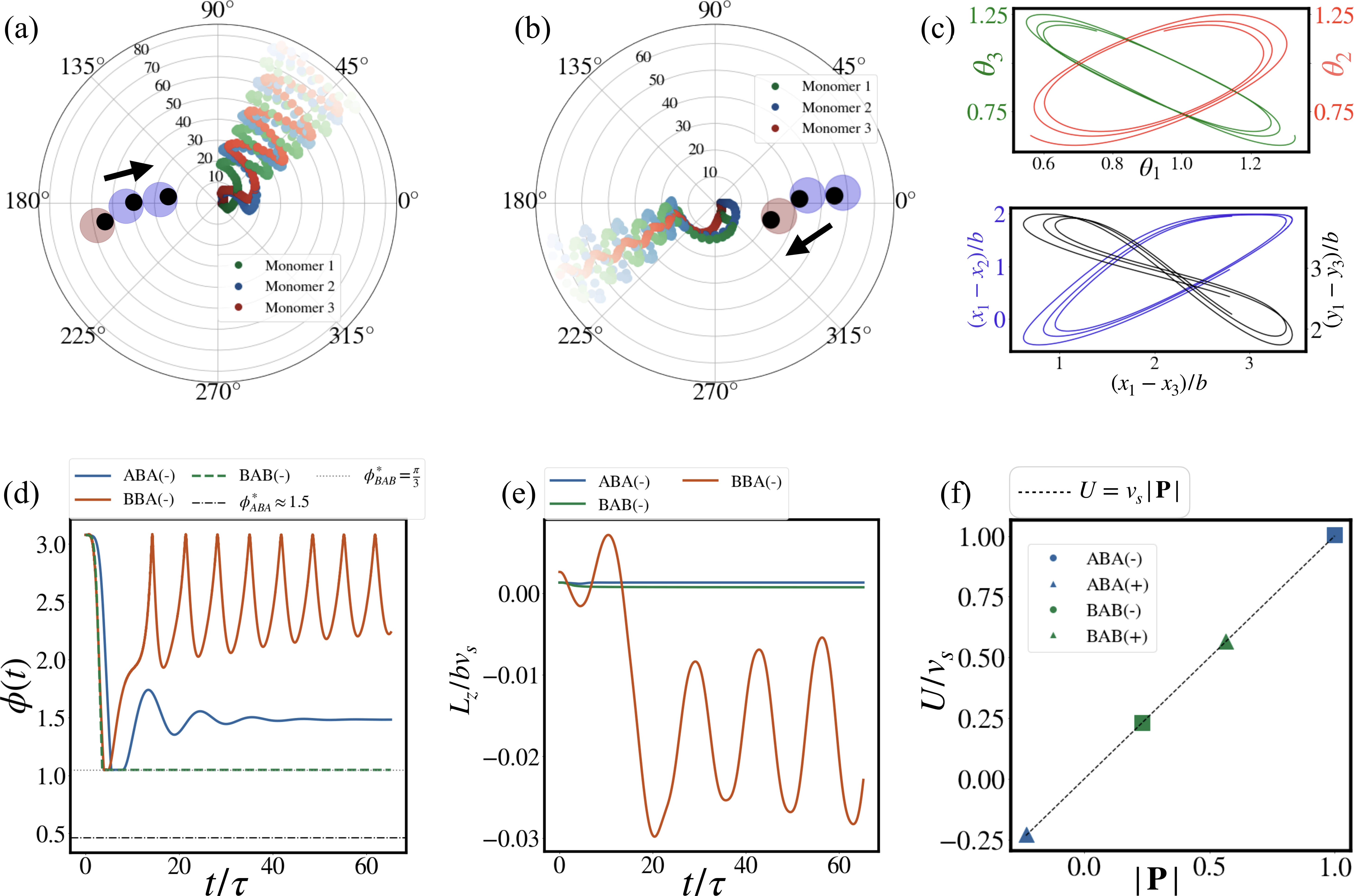}
    \caption{Self-sustained undulatory gaits of the BBA trimer due to anti-symmetric chemical affinities. Panel (a) and (b) display polar plots for the dynamics of BBA(-) and BBA(+) trimers respectively. Panel (c) displays the underlying attractor with the coupling of head and middle orientations to the tail (\textit{top}) and (hence) a coupling in the reduced coordinate space of $(x_{1}-x_{2},x_{1}-x_{3})$ and $(y_{1}-y_{2},x_{1}-x_{3})$ (for BBA(-)). Note that shared y-axes are used here, the respective quantities are colored same in the plot as in their axis labels. Panel (d) displays both $\phi(t)$, quantifying bond angle dynamics of configurations. Vertical dotted lines indicate steady-state $\phi$ attained by the respective configurations. Panel (e) $L_{z}$, capturing the undulation of BBA versus the other two.  In (f) validation of (\ref{eq:u_vs_pol}) is shown for ABA(-),BAB(-),ABA(+), and BAB(+) configurations (excluding the BBA configurations).
}
    \label{fig8}
\end{figure*}
%%%_---------------------
%%------------------------
The results are displayed Fig.~(\ref{fig7}) and Fig.~(\ref{fig8}).
% \textit{Top rows}(a)-(c) display the polar plots for the respective configurations. 
The dynamics of ABA(-) trimer is shown in Fig.~\ref{fig7}(a).
The ABA(-) trimer displays oscillatory angular dynamics, where the edge A monomers are repelled from the center and are equally repelled from each other. They eventually stabilize at an angle of $\phi \approx 105.9 ^{\circ}$, as shown in Fig.~\ref{fig8}(d) (blue curve). For ABA and BAB where the interactions are symmetric, $\phi$ can be shown to uniquely specify the dynamics, and that the effective system lies in $\{ \phi, \theta_{1} \}$. We get (see Appendix(\ref{app:dim_red_ABA})):
\begin{align}
    \dot{\phi} = \frac{v_{s}}{2b} \sin  \theta_{1} \, \sec\frac{\phi}{2}  \quad 0 \leq \phi < \pi
\label{eq:phi_aba}
\end{align}
Thus, the bond angle stabilizes when $\theta_{1} = 0$ as in the ABA case (Fig.~\ref{fig7}(a)).\\

Similar to the procedure outlined above, the dynamics of the ABA(-) can be reduced 
%$\{ x_{1},x_{2},y_{1},y_{2},e_{x_{1}},e_{x_{2}},e_{y_{1}},e_{y_{2}} \} \rightarrow \{ R_x\}$ at the \textit{fixed point} 
[see Appendix (\ref{app:dim_red_ABA})], 
\begin{align}
    \frac{dR_x}{dt} = v_{s}\left( 2 \cos(\theta_{1}^{*}) + \cos(\theta_{2}^{*}) \right) = 3v_{s}
\label{eq:ABA_red_fp}
\end{align}
thus the stable propelling trimer is analogous to Eq.~(\ref{eq:ab_red_fp}) with a proportional propulsion speed. 

The dynamics of BAB(-) configuration in shown in Fig.~\ref{fig7}(b). We see that the trimer \textit{reverses} its direction. The cardiod-resembling polar trajectory plateaus at $|R (t)| \approx 14 b$, after which the trimer reverse in direction, eventually reaching the origin. The mechanism for this is clear - the edge B monomers are repelled from the central A but are mutually attracted to one another. The equation governing the fixed point propulsion analogous to Eq.  (\ref{eq:ABA_red_fp}) is 
\begin{align}
    \frac{dR_x}{dt} = v_{s} \left(1 + 2\cos(\theta_{1}^{*}) \right)
\label{eq:BAB_red_fp}
\end{align}
Noting that $\theta_{1}^{*} \approx \pi$ (Fig. \ref{fig7}(b)), we see that the center of mass propels in the -ve $x$ direction. This ``reversal time scale" can be tuned by setting $\frac{\tau_f}{\tau}$ accordingly (for large $\frac{\tau_f}{\tau}$ the system will reverse slower). The steady state $\phi$ of the trimer is straightforwardly given by $\frac{\pi}{3}$ (close packed), which we see in Fig. \ref{fig7}). Movies for the ABA and BAB configuration are in SI Videos 8 and 9\cite{siText}. For the ABA(+) and BAB(+) configurations, the dynamics are more straightforward - simply stably propelling trimers (Fig. \ref{fig7}(c) and (d)).

Finally, we have the result for the BBA(-) configuration in Fig. \ref{fig8}(a), along with BBA(+) in Fig. \ref{fig8}(b) (see SI Video 10 \cite{siText}). We see an emergent undulatory gait motion, reminiscent of a micro-swimmer \cite{dreyfus2005microscopic}. The behaviour is driven by a competition for alignment between the tail A and the central B monomer with the leading B monomer. After an initial transient, the system settles into an oscillatory state, which lasts indefinitely within the time scale of the simulation. This oscillatory behaviour is characterised by limit cycles in phase space, as is shown in Fig.~\ref{fig8}(c). On the top panel, we see that both $\theta_{3}$ and $\theta_{1}$ are independently coupled to $\theta_{1}$. On the bottom panel, we see that, as a consequence, the relative coordinates $\mathbf{x}_{1}-\mathbf{x}_{2}$ and $\mathbf{x}_{1}-\mathbf{x}_{3}$ are coupled. The oscillations are captured by $\phi$ and $L_{z}$ (\ref{eq:ang_mom}) - see Fig. \ref{fig8}(d) and (e). In particular, $L_{z}$ differentiates BBA(-)'s dynamics with the other configurations, as undulations give rise to time-periodic oscillations in velocity displacements about the body molecule CM. 

Such temporal periodic waves have also been reported in other non-reciprocal models of active matter \cite{saha2020scalar, you2020nonreciprocity, scheibner2020odd, tan2022odd}, with the key difference being that in these the activity arises purely due to the non-reciprocity, with the systems otherwise being in equilibrium. Our travelling states (propagated via the BBA body) here are thus driven by \textit{both} the polar nature of the chain and the underlying anti-symmetric chemical interactions. These self-sustained gaits could also hint to a simple ``chemotactic gait" mechanism for microorganism swimming without the need to recourse to hydrodynamics \cite{ishimoto2023odd}, and/or mechanical models \cite{fang2010biomechanical}. Swimming gaits here arise purely due to competing chemotactic mechanism and are self-sustained via the non-linearities of the chemical trails in (\ref{eq:curr}).
% We note that there are two other configurations for the trimer case. The results for these two cases: ABA(+) and BAB(+) are shown in Appendix (\ref{app:trimer}). 
Finally, we also have a validity of our scaling relation (\ref{eq:u_vs_pol}) for the ABA and BAB configurations, owing to their symmetric histories of the trajectories (see Appendix (\ref{app:dim_red_ABA})). This is displayed in Fig. \ref{fig8}(f).

\section{Chiral active foldamers}
\label{sec:rotors}
\begin{figure}[b]
    \centering
    \includegraphics[width=.94\textwidth]{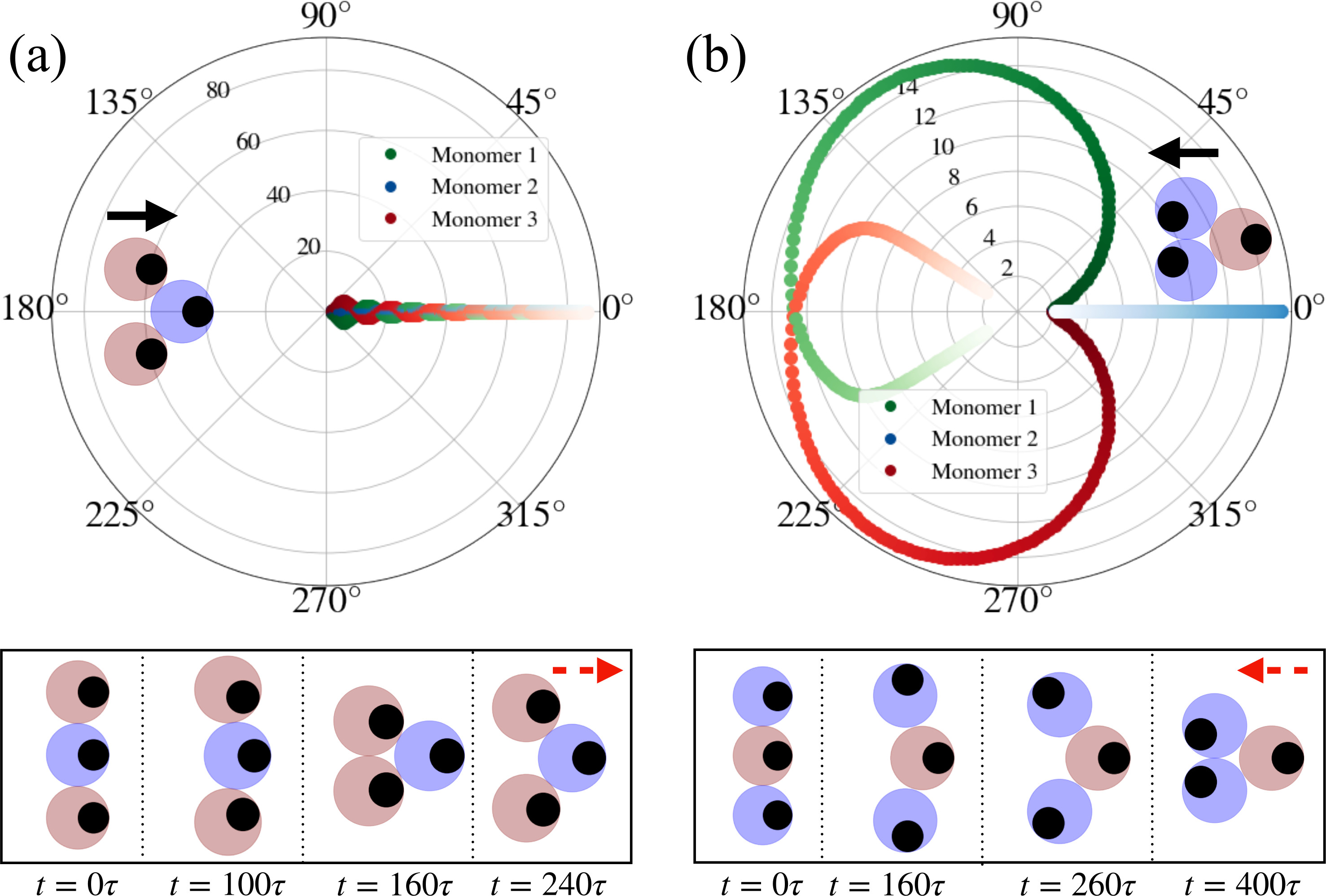}
    \caption{Different types of chiral rotors. Panel (a) illustrates the stable rotations of the $B_7 A_4$ molecule. Panels (b) and (c) illustrates the rotational dynamics for example cases of Foldamer F1a ($A_6 B_6 A_6$ configuration) and Foldamer F1b ($A_2 B_{6} A_2$ configuration) respectively. These spinning molecules have free arm segments, that act as rotor blades. Panel (d) illustrates the Foldamer F2 rotational dynamics, characterized by attached arm segments, for the $A_4 B_{10} A_4$ configuration. For all panels initial configuration is shown on the left, with $\Upsilon$ matrix entries in bottom right. Red curved arrows indicate instantaneous change propulsion direction, whilst black curved arrows indicates rotation about CM. For panels (b) and (c), $k_{sp} = 3.48 \times 10^{4} \mu m / s^{2}$ and for panel (d), $k_{sp} = 1.74 \times 10^{5} \mu m / s^{2}$. For all plots and insets, we use reddish color for A monomers and bluish color for B monomers. Color shading is to specify initial position along the chain.
}
    \label{fig9}
\end{figure}

\begin{figure}[t]
    \centering
    \includegraphics[width=.4\textwidth]{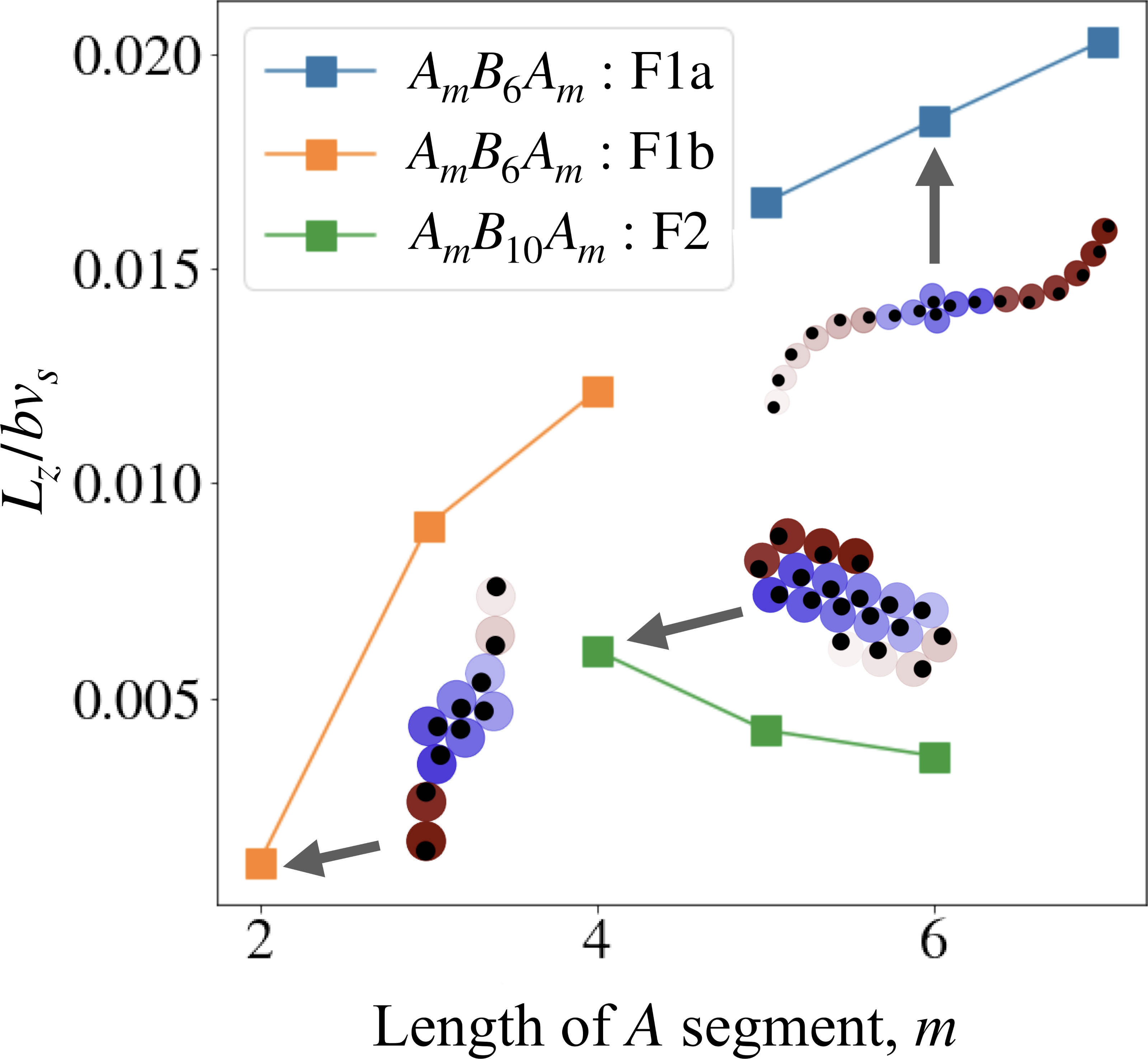}
    \caption{Tunable rotational frequency of foldamers. Rotational frequencies (non-dimensionalized) of Foldamers F1 ($A_m B_6 A_m$) and F2 ($A_m B_{10} A_m$), ${L_z  }/{v_sb}$, are plotted versus the length of the A type segments, $m$. Foldamers F1a and F1b are shown in blue and yellow, with Foldamer F2 in green. Insets show molecule corresponding to the $m$ value indicated by the black solid arrow.
    Color scheme is same as in Fig.\ref{fig9}.
}
    \label{fig10}
\end{figure}
The trimer results above hint to a broader design principle for active colloidal molecules: the ability to acquire specific steady-state structure and/or dynamics by carefully designing the interactions and/or initial conditions. We extend on this idea in this section, where we show that we are able to construct molecules with a \textit{chirality} in the dynamics. For this section, we term our final collapsed structure as an \textit{active foldamer} \cite{zeravcic2014size, zeravcic2017colloquium}. The brief outline of the mechanism is as follows. For a monodisperse chain with a single segment, the foldamer consists of a single $\mathbf{P}$ vector, passing through its center of mass. For foldamers of more than one species, each distinct segment of the chain has it's respective $\mathbf{P}^{(\alpha)}$, but now each of the segment's $\mathbf{P}^{(\alpha)}$ no longer passes through the CM of the foldamer. As a result, rotations are induced which are proportional to the (perpendicular) distance of the  $\mathbf{P}^{(\alpha)}$ vectors to the CM. For the rest of the segment we denote the segment indices as $\{ \alpha_1, \alpha_2, ..., \alpha_M \}$ for $M$ generic segments. This mechanism can produce two phenomenon: (i) planar rotation about a point in the case where the composite molecule has a non-zero net $\mathbf{P}$, and (ii) rigid body rotations about the molecule CM (``spinning"), if the net $\mathbf{P}$ is (designed to be) zero. A generalized form of Eq. (\ref{eq:ang_mom}) gives the rotational frequency of the molecules $\frac{L_z}{b^{2}}$, with here $L_z = v_{s}\sum_{\alpha}^{M} \Big( \mathbf{P}^{\alpha} \times (\mathbf{r}^{\alpha} - \mathbf{R}) \Big)$.
An example of the former is shown in Fig. \ref{fig9} (a), which is a stable (elliptical) orbit for the $B_{7}A_{4}$ configuration (see SI Video 11 \cite{siText}). Here, there is a ``nucleus" of B type being propelled by a ``rim" of A type. The net handedness is clockwise. These dynamics are stable, in that the positions and orientations of the rotor do not change in the steady-state - i.e there is a fixed chirality in the system that stably persists. The inset of Fig. \ref{fig9}(a) shows the initial conditions of the $B_7 A_4$ configuration; first two segments are identical to (\ref{eq:init_cond}), with $\theta_{i}^{\alpha_{3}} = \pi \quad \forall i$.

Constructing instead a foldamer that has a net zero polarization, but $\mathbf{P}^{\alpha}$ segments symmetrically apart from the CM in opposite orientations, one can design dynamics that exhibit rotations about the foldamer CM - i.e. a spin degree of freedom. The simplest examples of these are molecules with symmetrically located oppositely aligned $\mathbf{P}^{\alpha_{1}}$ and $\mathbf{P}^{\alpha_{3}}$, along with a central segment of $\mathbf{P}^{\alpha_{2}}=0$. Attaining such structures require a judicious choice of $\Upsilon_{BB}$, $\Upsilon_{AB}$, and $\Upsilon_{BA}$ entries; examples of these are indicated in the insets of Fig. \ref{fig9}. 
We report here two such foldamers, distinguished by the position of the ``arm" segments with respect to the central body - i.e whether they are free or attached to the central body. We use the label Foldamer F1 for those that have ``free-arm" segments of species A. The final collapsed structure is highly sensitive to the initial conditions. Labelling the length of the A-type arm segments as $m$, we obtain different structures for Foldamer F1 with variation in $m$ - we label this with the prefixes a/b to distinguish the two Foldamer F1 structures. An example of Foldamer F1a is the $A_6 B_6 A_6$ configuration shown in Fig. \ref{fig9}(b). An example of Foldamer F1b is the  $A_2 B_6 A_2$ configuration is shown in Fig. \ref{fig9}(c), with the initial conditions and $\mathbf{\Upsilon}$ matrix shown in the inset (\textit{left} and \textit{bottom right}). For Foldamer F2, the A-type arms are attached onto the B-type central body; an example of this is shown in Fig. \ref{fig9}(d) for the $A_4 B_{10} A_4$ configuration (respective $\mathbf{\Upsilon}$ matrix again found in inset). 

The rotational frequency of these foldamers are thus straightforwardly seen to be determined by the contribution of the arm segments' $\mathbf{P}^{\alpha}$, and that they are \textit{tunable} via variation of the length of the arm segments. This is shown in Fig. \ref{fig10}, with the rotational frequency ${L_z}/(bv_s)$ plotted against the length of the A arm segment, $m$. Foldamers F1 and F2 display different response in their rotational frequency to an increase in $m$, namely Foldamers F1 enhance their rotational frequency whilst Foldamers F2 \textit{decrease} their frequency. For Foldamers F1, this occurs since $\mathbf{r}^{\alpha} - \mathbf{R}$ increases with $m$ (in addition to the commensurate increase in $\mathbf{P}^{\alpha}$ - see Fig (\ref{fig_cr})). Movies of Foldamers F1a and F1b are shown in SI Videos 12 and 13 respectively \cite{siText}.
For Foldamers F2, instead $\mathbf{r}^{\alpha} - \mathbf{R}$ decreases with $m$. We note that the range of $m$ probed here instead is relatively narrow, only $m \in \{ 4,5,6 \}$, since for $m<4$ and $m>6$ the structures change significantly (not shown in this paper). These indeed could be other classes of Foldamers by themselves, but we do not pursue the study of them any further. Movie of Foldamer F2 can be seen in SI Video 14 \cite{siText}. A longer range of $m$ to display the tuning could in principle be done with a foldamer with a longer B body segment - we leave this to future work.

We have thus shown in this section that our model produces two distinct types of chirality in active foldamer dynamics, with the latter being tunable. The tunability can also be positive (increasing with segmential length) or negative depending on whether the foldamer has free or attached arm segments.
 We clarify here that whilst the segmential polarization mechanism here is generically applicable for any  chemical affinity matrix between the distinct species, the spinning rotors that we observe require a \textit{non-reciprocity} in the species interactions (e.g. any composite molecule with arbitrary $\mathbf{\Upsilon}$ will have dynamics that break handedness symmetry, but they may not acquire a spin degree of freedom). 

\section{Summary and discussions}\label{sec:summary}
In summary, we have shown how a simple model of a chemically self-interacting colloidal chain can support a wide range of dynamical phenomenology, including rigid flocks, self-sustained undulatory gaits and chiral rotors. In particular, we have uncovered (i) the role of chemical trails in providing an sustained asymmetry in the dynamics, and (ii) the role of non-reciprocal (internal) alignment interactions in the collective dynamics, which in addition we have shown to be tunable.
The former necessitates two particular phenomena reported here, firstly, the universal C-shape rigidity in the chemo-repulsive chain. Here, the historical build-up of chemicals in (\ref{eq:curr}) renders a forward-backward asymmetry in the dynamics of the chain, resulting in a universal C-shape flock (see SI Video 6 \cite{siText}). Without such an asymmetry, the C-shape acquisition ceases to be universal, and dependent solely on the initial conditions. 
Second, it gives rise to a self-sustained \textit{chemotactic undulations} of a binary species colloidal molecule with tail, body, and head monomers ~\cite{dreyfus2005microscopic, nishiguchi2018flagellar}, that self-sustains its microswimmer gait without any hydrodynamics and/or biomechanics. The tail-head A-B alignment interaction is mediated via the respective collective memory of the past trajectories, which is necessary for the undulatory gait to sustain itself. These results thus also point towards a novel function of trail-dependence in the collective dynamics in active matter \cite{sengupta2009dynamics, taktikos2012collective, hokmabad2022chemotactic}, namely that they sustain a \textit{spatio-temporal} asymmetry in the dynamics (i.e. particular pattern may be universally selected due to the shared collective memory of the past) - which in our study manifests as a global polar order of the chain. The effect of hydrodynamics on this function remains an open question.  

The results for our binary species chain with non-reciprocal chemical affinities (oscillations, reversal of direction of motion, undulatory gaits, chirality) hint towards design principles of active molecules, where the ``structure-function" paradigm of self-assembly holds; an example of which we have reported are chiral molecules. Various other ``structure-function" specific molecules are possible, however; for instance in SI Video 15 \cite{siText} we report the ``nucleus + envelope" molecule, where a stable nucleus of species B is shielded with a rim of species A, with the overall molecule propelling stably in the positive $x$-direction. Extending this to multiple species, we envisage that our model can be used to program active dynamical features of colloidal molecules \cite{zeravcic2017colloquium} via these aforementioned parameters - interaction types, initial orientations, and segmental lengths; this will be pursued in a future work. {Experimentally, we note that chasing interactions in colloids have been realized (\emph{e.g.} in \cite{meredith2020predator}), and our results thus provide testable predictions for phenomenon described here.}

Chirality, as shown here, can exist in any generic multi-segment configuration, whilst a spin degree of freedom (rotations about the body CM) can be incorporated by a chemo-attractive body segment, and chemo-repulsive arm segments that are oppositely aligned. Connecting to biological systems, we note that chiral dynamics is a ubiquitously reported feature - for instance in MDCK cell assemblies \cite{segerer2015emergence, chin2018epithelial} and the actomyosin cortex-cell nucleus complex \cite{kumar2014actomyosin}. It is interesting to note that our model provides a clear mechanism for the chirality via the internal non-reciprocal nature of the chemical affinities in the dynamics of the chain, without resorting to any hydrodynamics \cite{kumar2014actomyosin, singh2016universal, tan2022odd}. The tunable nature of the rotational frequency of such molecules further suggest that such structures can be used as microscopic clocks; it would be interesting to see if biological systems apply similar time-keeping mechanisms for their function. 

We note here that stable C-shape chains from chemo-repulsive self-interactions were first reported in \cite{kumar2023emergent}. Here, we have done a detailed analysis of this structure and further studied the phase behaviour of the chemo-repulsive case. {In particular, we have shown the the relation of Eq.\eqref{eq:u_vs_pol} holds for the speed of the chain and explained its origin.} The results reported for the chemo-attractive and chains with binary species {with non-reciprocal chemical interactions} have been reported here for the first time. In particular, the phenomenology reported in the latter case - self-generated direction reversal, undulatory gaits, and chiral foldamers - are reported here for the first time, to the best of our knowledge. 

{In this work, we have studied the dynamics of a single polymer with chemical interactions in 2D. Extensions of this work could be the study of many interacting chains, where the effects of topological defects are imminent, as well as to 3D. In terms of connection to experiments, we note that at a modelling level, we have ignored hydrodynamic interactions. This is rationalised by the fact hydrodynamic fields decay more quickly than chemical fields in strong 2D confinement in Hele-Shaw cells, and thus do not affect our results significantly~\cite{kumar2023emergent}. On the other hand, fluid flow can significantly affect dynamics in studies of 3D systems \cite{bechinger2016}, and thus, need to be included. These suggest exciting avenues for future work. }
%Experimental design of chasing interactions in colloids have been realized (\emph{e.g.} in \cite{meredith2020predator}), and our results thus provide testable predictions for phenomenon described here.\\\par

\section*{Acknowledgements}
We thank Ronojoy Adhikari, Mike Cates, K. Vijay Kumar,  Ignacio Pagonabarraga, and John Toner for useful comments. AGS acknowledges funding from the DIA Fellowship from the Government of India. ST thanks the Department of Atomic Energy (India), under project no.\,RTI4006, the Simons Foundation (Grant No.\,287975), the Human Frontier Science Program and the Max Planck Society through a Max-Planck-Partner-Group for their support. RS acknowledges support from seed and initiation grants from IIT Madras as well as a Start-up Research Grant from SERB (SERB File Number: SRG/2022/000682), India.

% \newpage
\appendix
% \widetext
\section{Simulation details}\label{app:simDetailsParams}
We simulate Eq.~(\ref{eq:dyn}) using a forward Euler-Maruyama method to evolve the positions and orientations of the particles in time. 
%with the parameters listed in Appendix \ref{app:simDetailsParams}, unless otherwise stated. 
For the entire paper (unless specified otherwise) we use the following initial conditions:
% \begin{subequations}
\begin{align}
\label{eq:init_cond}
    \theta_{i}(0) = 0, \qquad x_{i} (0)= 0, \qquad \forall i,
    \qquad\qquad 
    y_{i+1}(0) = y_{i}(0) + 2b, \qquad\, 1 \leq i \leq N-1
\end{align}
% \end{subequations}
Here $\theta_i$ is the angle made by the orientation $\mathbf e_i$ of the 
$i$th particle with the $x$-axis. 
Thus, Eq.\eqref{eq:init_cond} implies that the initial chain is aligned vertically (upward facing in the plane of the page in this paper). 
%In addition, we have observed that the results are universal and do not depend on the initial orientations (see SI Video 6\cite{siText}). A detailed study of the effect of initial conditions is not performed in this work.
Variation of translational noise is naturally important when considering the statistical mechanics of long `conventional polymer(s)', whose constituents have length dimensions $\sim 100 nm$; here we instead study the precise \textit{dynamics} of these chains. Indeed, using experimentally feasible quantities of $b \sim 25~\mu$m, viscosity of solvent $\eta \sim 10^{-3}$ \cite{kumar2023emergent}, we find that the typical active force [$\mathcal {O}(6\pi\eta b v_s) \sim 10^{-11} N$], dominates the typical Brownian forces [$\mathcal O(k_BT/b)\sim 10^{-16}N$], and thus, $D_{t}$ is ignored in this paper. $D_{r}$ on the other hand,
plays a crucial role in the phase determination of the chain, as shown in the paper.

The model described in Section \ref{sec:2} has been numerically simulated in an unbounded two-dimensional (2D) space using a custom Cython code (Python on top of a C backend). Throughout the paper, we have used: $b=25\mu m$, $v_{s}=50\mu m/s$, $k_{sp}=1.74 \times 10^{4}\mu m /s^{2}$, $k_{e}=8.70 \times 10^{4}\mu m /s^{2}$, $c_0 = 1$ and $\chi_{r} = 7.5 \times 10^{3} \mu m^{3} /s^{2}$. For the chemo-repulsive case (Figure \ref{fig_cr}), $dt=0.01$. 
For the chemo-attractive and binary species cases 
(Figure \ref{fig_ca}, \ref{fig6}, \ref{fig7}, \ref{fig8}) $dt=0.0018$. Unless varied (as in Figures \ref{fig_cr}, \ref{fig_ca}) $D_{c} = 2500 \mu m^{2} /s$, $\mathcal{A}_{1} = 10^{-2}$, $\mathcal{A}_{2} = 0$ (with $D_{r} = 0$). For all sections $\frac{\chi_{t}}{\chi_{r}} = 10^{-4}$, with the exception of Section \ref{sec:rotors} (Fig \ref{fig9}), where $\frac{\chi_{t}}{\chi_{r}} = 10^{-2}$. In
Fig. \ref{fig_cr}(b), the parameters used for 
$(\mathcal{A}_{1},\mathcal{A}_{2})$ are (i) $(10^{-1},10^{-1})$, (ii) $(10^{-1},10^{1})$, (iii) $(10^{1},10^{3})$, (iv) $(10^{-1},10^{4})$. 
In Fig.\ref{fig_ca}(b), the parameters used for $(\mathcal{A}_{1},\mathcal{A}_{2})$ are (i) $(10^{-1},10^{-1})$, (ii) $(\mathcal{A}_{1},\mathcal{A}_{2})$ are (i) $(10^{-1},10^{1})$, (iii) $(10^{1},10^{3})$, (iv) $(10^{-1},10^{4})$.
%%
%%
%%---------------------------------------------------
%%---------------------------------------------------
\subsection{Curvature of the polymer}\label{app:curv}
%%---------------------------------------------------
%%---------------------------------------------------
 We quantify the curvature of the chain via the Monge representation \cite{helfrich1973elastic}. In particular, we compute the curvature as:
\begin{align}
  \kappa = \left\langle  \nabla \cdot \left[ {\nabla h_{i}} / {\sqrt{1 + | \nabla h_{i} |^{2}}} \right]  \right\rangle,
% $%
\end{align}
where $h_{i}$ is the height function of the chain, evaluated at each monomer location, measured from the vertical line connecting the edge monomers. The average is performed across the monomers in the chain, giving a scalar value. In our case, the inner bracket is simply reduced to $\partial_{y} \left[ {\partial_{y} x_i(y)}/{\sqrt{1 + | \partial_{y} x_i |^{2}}} \right]$.
{The curvature of the chain is the order parameter used to distinguish different steady-state configurations for an active
polymer with chemo-repulsive monomers  (see section \ref{sec:chemo-rep}).
For Fig. \ref{fig_cr}(a), we distinguish different phases by taking the averaged $|\kappa|$ in the steady state. The values used to distinguish the different phases are: (I+) $|\kappa| \leq 0.05$, (II+) $0.05 < |\kappa| \leq 0.1$. (III+)$|\kappa| \leq 0.05$ (identical to Phase I+ but sufficiently separated), (IV+) $|\kappa| > 0.10$. Thus, the C-shape is identified by the highest time averaged curvature.} 
%%---------------------------------------------------
%%---------------------------------------------------
\subsection{Coordination number}\label{app:coordZ}
%%---------------------------------------------------
%%---------------------------------------------------
 We now define a \textit{coordination number} ($z$) for each monomer
\begin{align}
    z = \{ | \mathcal{N}_{i} |, \forall i \}, \qquad \mathcal{N} = \{ |\mathbf{r}_{i} - \mathbf{r}_{j}| < 2b, \forall j \neq i \}
\end{align}
with the $\max (z)$ being an order parameter used to distinguish different steady-state configurations for an active polymer with chemo-attractive monomers (see section \ref{sec:chemo-attract}).
{For Fig. \ref{fig_ca}(a), these are now distinguished by the averaged maximum coordination number along the chain, $\max (z)$, averaged over the steady-state. Hexagonally packed structures have $\max(z)=6$, whilst structures with no collapse at all have $\max(z)=2$. The values used to distinguish the different phases are: (I-) $\max(z) = 2$, (II-) $2 < \max(z) \leq 6$. (III-) $\max(z) = 2$ (identical to Phase I- but sufficiently separated), (IV-) $\max(z) = 2$.}

\section{Dimensionality reduction from the symmetries of the dynamical system}\label{sec:SymmSec}
\begin{figure} 
    \centering
    \includegraphics[width=.8840\textwidth]{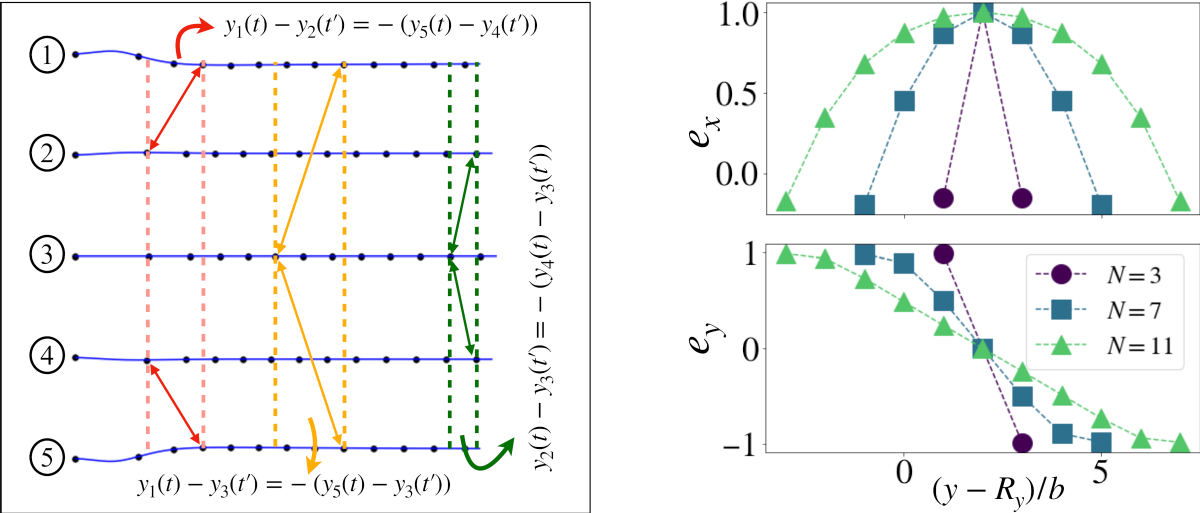}
    \caption{Illustration of the symmetry of the dynamics for the chemo-repulsive case of $N=5$. LEFT PANEL: trajectory of each monomer in the chain is shown (labelled on the left), with time points colored in black. The equalities in (\ref{eq:symm1}) can be deduced by taking vertical distances at different time points (a particular $(t,t')$ pair is uniquely coloured). Equalities of (\ref{eq:symm2}) is deduced by considering the time spacings. RIGHT PANEL: steady-state orientations ($e_x$ and $e_y$) satisfying Eq.~(\ref{eq:fpoints_even}) are shown, for $N=3$, $N=7$, and $N=11$. Dotted lines are a guide to the eye.
}
    \label{fig11:symm}
\end{figure}

To write down reduced dynamical equations for dimers and trimers, one has to reduce the dimensionality of the phase space. In this section, we will outline the dimensionality reduction procedure that gives the generic relation Eq. \eqref{eq:u_vs_pol}, along with reduced dynamics of \eqref{eq:ab_red_fp}, \eqref{eq:ABA_red_fp}, \eqref{eq:BAB_red_fp}.
To prove Eq.\eqref{eq:u_vs_pol}, as an instructive case, let us consider the time trajectory of the $N=5$ chain, shown in Fig.~(\ref{fig11:symm})(a). The dynamics can be seen to be fully symmetric about the central monomer, and the following equalities can be deduced
%\begin{subequations}
\begin{align}
\label{eq:symm1}
    y_{1}(t) - y_{2}(t') = y_{4}(t')-y_{5}(t), 
    \qquad
    y_{1}(t) -  y_{3}(t') = y_{3}(t') - y_{5}(t), 
    \qquad
    y_{2}(t) -  y_{3}(t') &= y_{5}(t') - y_{4}(t) 
\end{align}    
%\end{subequations}
and further
\begin{align}
    x_{1}(t) = x_{5}(t), \quad x_{2}(t) = x_{4}(t), \quad \forall t
\label{eq:symm2}
\end{align}

This symmetry, as we show in Appendix (\ref{app:orient_fp}), allows us to show that the fixed points of the orientations are symmetric in one and anti-symmetric in another direction:
\begin{align}
\label{eq:fpoints5}
    e_{x_{1}} = e_{x_{5}}, \quad e_{y_{1}} = -e_{y_{5}},
    \quad
    e_{x_{2}} = e_{x_{4}}, \quad e_{y_{2}} = -e_{y_{4}},
    \quad
    e_{x_{3}} = 1,\qquad  e_{y_{3}} = 0
\end{align}
% \end{equation}
For an arbitrary chain of $N$ monomers, this generalizes to the condition:
\begin{align}
    e_{x_{i}} = e_{x_{N-i+1}}, \quad e_{y_{i}} = -e_{y_{N-i+1}}, \quad \forall i \in \{ 1,2,...,N \}
\label{eq:fpoints_even}
\end{align}
Thus, a spatial symmetry in the problem implies an orientational symmetry.
%The above symmetry also implies $\theta_i = \theta_{N-i+1}$.  
Note also that for odd-numbered chains, (\ref{eq:fpoints_even}) implies that $(e_{x},e_{y})=(1,0)$ %, or the condition that $\theta=0$, 
for the monomers in the (arithmetic) center.
This symmetry is shown in Fig.~(\ref{fig11:symm}b), where the orientations are plotted in the CM frame perpendicular to the propagation direction, in the continuum parametrization, showing that $e_{x}(y-R_y) = e_{x}(-(y-R_y))$, (\textit{top panel}) and $e_{y}(y-R_y) = -e_{y}(-(y-R_y))$ (\textit{bottom panel}) (the relation between the two given straightforwardly by $e_{y} = \sqrt{1 - e_{x}^{2}}$). Note that this straightforwardly implies that $\theta(-y) = -\theta(-(y-R_y))$. Summing up Eq.~(\ref{eq:dyn}) across the chain, and using Eq.~(\ref{eq:pol}), we obtain Eq.~(\ref{eq:u_vs_pol}). 

We will show, in this and the following sections, that the scaling of Eq.~(\ref{eq:u_vs_pol})
holds true \textit{whenever} the symmetries in (\ref{eq:symm1}) and (\ref{eq:symm2}) are applicable (for instance during a certain duration of the dynamics). We will also show in Sections (\ref{sec:chemo-attract-rep}) that these symmetries can reduce the dimensionality of the system for dimer and (selected) trimer configurations. We also note that this scaling also holds true for arbitrary initial conditions not given by Eq.~(\ref{eq:init_cond}) \cite{kumar2023emergent}. In this case, (\ref{eq:u_vs_pol}) does not apply throughout the dynamics, but merely in the \textit{steady-state}, when the chain acquires the C shape (irrespective of the initial condition) that characterizes in phase IV+.

\subsection{Derivation of orientational fixed point symmetry}
\label{app:orient_fp}

% \begin{figure} 
%     \centering
%     \includegraphics[width=.4\textwidth]{fig11.png}
%     \caption{Schematic of $N=5$ motion, exhibiting the symmetry of positional degrees of freedom. Centers of monomers indicated by black dots, dotted lines indicating surfaces. To be complemented with Fig.~(\ref{fig11:symm}).}
%     \label{fig_app0}
% \end{figure}
We derive here the condition for the orientational fixed points for a representative $N=5$ case, given by (\ref{eq:fpoints5}). This is shown in Fig.~(\ref{fig11:symm}). 
From (\ref{eq:dyn2}), note that the orientational fixed points are reached if $\mathbf{e}_{i}^{*} \propto \mathbf{J}_{i}$. Let us define
$\Delta x_{ij}(t,t') = x_{i}(t)-x_{j}(t')$ and 
$\mathcal{F}_{ij}^{(x)}(t,t') = 
\exp\left[-\frac{[\Delta x_{ij}]^{2}}{\mathcal{D}(t,t') } \right] 
$
and likewise for $\Delta y_{ij}(t,t')$ and $\mathcal{F}_{ij}(t,t')^{(y)}$ by the substitution $x \rightarrow y$. Then, using (\ref{eq:curr}), we have:
\begin{equation}
    e_{x_{i}}^{*} \propto \int dt' \sum^N_{\substack{j=1
    \\i \neq j    }}
    \left[ \frac{\Delta x_{ij}(t,t')}{\mathcal{D}(t,t')}\mathcal{F}^{(x)}_{ij}(t,t')\right].
\label{eq:app_fp_integrals}
\end{equation}
The symmetries given by (\ref{eq:symm2}) renders $\Delta x_{1i} = \Delta x_{5i}$, $\Delta x_{2i} = \Delta x_{4i}$,  $\forall i$. 
Thus, we thus see straightforwardly that $e_{x_{1}}^{*}=e_{x_{5}}^{*}$ since $\Delta x_{12} = \Delta x_{52}$, $\Delta x_{13} = \Delta x_{53}$, $\Delta x_{14} = \Delta x_{54}$, and $\Delta x_{11} = \Delta x_{55}$. Now using $\Delta x_{22} = \Delta x_{44}$, $\Delta x_{23} = \Delta x_{43}$, $\Delta x_{24} = \Delta x_{24}$, and $\Delta x_{21} = \Delta x_{41}$, we have $e_{x_{2}}^{*}=e_{x_{4}}^{*}$. Since this is an odd-numbered chain $e_{x_{3}}$ takes a standalone value. We can show this to be $1$ by finding $e_{y_{3}}^{*}$, instead of explicitly evaluating the integrals. This proves the $x$ component of (\ref{eq:fpoints5}). For the $y$ component, we can repeat the procedure with $x \rightarrow y$ substitution in (\ref{eq:app_fp_integrals}). Using (\ref{eq:symm1}), we get
\begin{align}
\Delta y_{12} &= +\Delta y_{54}, \quad 
\Delta y_{13} = +\Delta y_{53},\quad
\Delta y_{14} =  - \Delta y_{52}, \quad 
\Delta y_{14} =  - \Delta y_{51} 
    \\
\Delta y_{21} &= -\Delta y_{45}, \quad  
\Delta y_{23} = -\Delta y_{43} 
\quad
\Delta y_{24} = -\Delta y_{42}, \quad 
\Delta y_{25} = -\Delta y_{41}
\label{eq:app_delta_y}
\end{align}
which implies that $e_{y_{1}} = -e_{y_{5}}$, $e_{y_{2}} = -e_{y_{4}}$ from (\ref{eq:app_fp_integrals}). To prove that $e_{y_{3}} = 0$, we use $\Delta y_{31} = -\Delta y_{53}$ and $\Delta y_{23} = -\Delta y_{32} $, which completes the proof of (\ref{eq:fpoints5}). %($e_{y_{3}} = 0 \implies e_{x_{3}} = 1$) 
For an arbitrary $N$, these results generalize straightforwardly to (\ref{eq:fpoints_even}). For odd numbered chains, the central monomer always has $e_{x} = 1$ and $e_{y} = 0$.

\subsection{Metastable crystallites' symmetric histories}
\label{app:metastable}
\begin{figure}%[b]
    \centering
    \includegraphics[width=.84\textwidth]{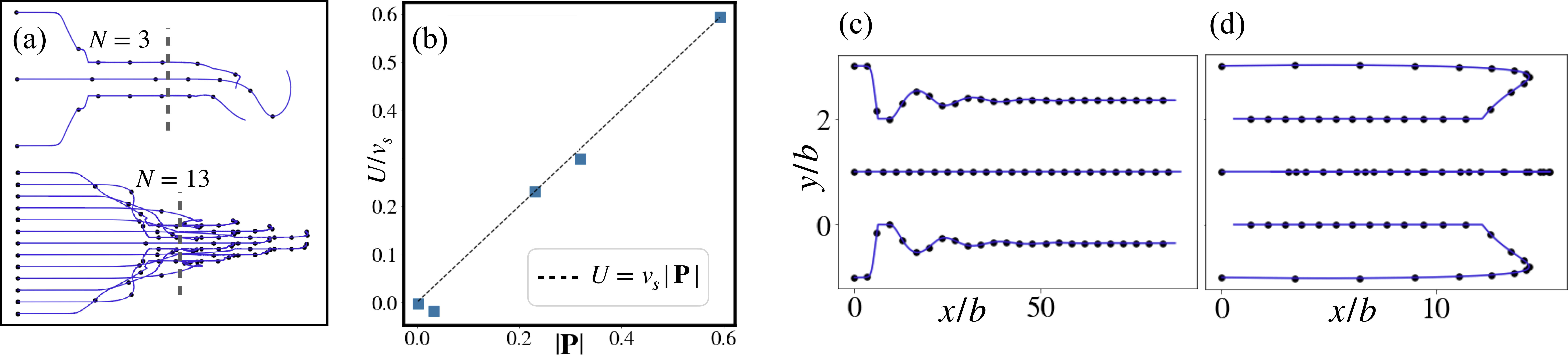}
    \caption{
    %Metastable crystallites in the chemo-attractive case, demonstrating range of validity of the symmetric-trails approach. 
    (a): \emph{top} shows that the symmetry of the trails holds for the metastable transient up to the collapse time (vertical dotted line), thus the averaging in (\ref{eq:u_vs_pol}) is performed for time points in the metastable state suitably before. On the bottom panel, we show that, even without a metastable lattice, hexagonally packed lattices also display such a symmetry. (b): plot of $U$ vs. $|\mathbf{P} |$ is shown for the metastable lattices ($N=3$,$4$,$5$,$6$, and $7$), showing a good agreement with (\ref{eq:u_vs_pol}). Panels (c) and (d) show symmetric trails of ABA(-), BAB(-) configurations.}
    \label{fig12_m}
\end{figure}

%%------------------------
Here, we analyse the metastable crystallites. Let us call the time at which the metastable crystallites lose their stability as the \textit{collapse time}. 
We can again exploit the symmetric histories argument, which we find to be applicable up to the collapse time. This is displayed in Fig. (\ref{fig12_m}a) (\textit{top row}). 
A straightforward consequence of this is the applicability of relation Eq.~(\ref{eq:u_vs_pol}) in this regime. 
This is displayed in Fig.(\ref{fig12_m})b, where the average in (\ref{eq:u_vs_pol}) is performed for time points before the collapse time. 
Data points here correspond to $N=3,4,5,6,7$ simulations. Beyond this regime, relation (\ref{eq:u_vs_pol}) is indeed no longer applicable. We note that for hexagonally packed crystals, $|\mathbf P | \sim 10^{-1}$ in this scaling regime, thus we do not display the data points in Fig. (\ref{fig12_m})b, though agreement with (\ref{eq:u_vs_pol}) also holds. 

%%=========================================================================
\subsection{Derivation of dimensionality reduction for monodisperse dimers}
\label{app:dim_red_mono}
To reduce the dimensionality of the dimer chain, using (\ref{eq:fpoints_even}), we have:
\begin{align}
    \frac{dR_x}{dt} &= v_{s}\left( \cos(\theta_{1}) + \cos(\theta_{2}) \right),\qquad
    \frac{dR_y}{dt} = v_{s}\left( \sin(\theta_{1}) + \sin(\theta_{2}) \right) = 0, \\
    \frac{d\theta_{1}}{dt} &= \chi_{r} (e_{x_{1}}\mathcal{I}_{b} - e_{y_{1}}\mathcal{I}_{11}^{(R_x)}), \qquad
    \frac{d\theta_{2}}{dt} = \chi_{r} (e_{x_{2}}\mathcal{I}_{b} - e_{y_{2}}\mathcal{I}_{22}^{(R_x)}), 
    %= (e_{x_{2}}\mathcal{I}_{b} - e_{y_{2}}\mathcal{I}_{11}^{(R_x)})
\label{eq:app_aabb}
\end{align}
where we have defined
\begin{align}
    \mathcal{I}_{b} &= \int_{0}^{t-1} \frac{2b}{(4D_{c}(t-t'))^{2}} \exp\Big( \frac{-4b^{2}}{\mathcal{D}(t,t')}\Big) dt',
    \qquad
    \mathcal{I}_{11}^{(R_x)} = \frac{1}{2} \int_{0}^{t-1} \frac{\Delta R_x(t,t')}{(4D_{c}(t-t'))^{2}} \exp\Big( \frac{-\Delta R_x^{2}}{4\mathcal{D}(t,t')}\Big) dt' = \mathcal{I}_{22}^{(x)}. 
\label{eq:app_integrals}
\end{align}
In the above, we have used $x_{1} = \frac{1}{2} R_x$ by the symmetry, and $\Delta x_{11}(t,t') = \Delta x_{22}(t,t')$. Thus, we see that the equations span a 3 dimensional phase space in $\{ R_{x}, \theta_{1}, \theta_{2} \}$.
To evaluate the integrals (\ref{eq:app_integrals}), we use the fact that $\frac{b}{l_{d}} \ll 1$, which applies in Phase IV, to expand the integrand $\exp\Big[ -\frac{(R_{x}(t) - R_{x}(t'))^{2}}{16D_{c} (t-t')} \Big] \approx 1 - O(\frac{b}{l_{d}}^{2})$. Thus
$\mathcal{I}_b = \int_{0}^{t-1} dt^{'} \frac{2b}{(4D_{c}(t-t'))^{2}} \Big[1 + O\left(\frac{b}{l_{d}}^{2}\right) \Big] \approx \frac{b}{16D_{c}^{2}}$. 
Similarly, for $\mathcal{I}_{11}$, we  have:
\begin{equation}
\begin{aligned}
    \mathcal{I}_{11} \approx \frac{1}{2} \dot{R}_{x} \int_{0}^{t-1} dt^{'} \frac{1}{(4D_{c})^{2}(t-t')} \Big[1 + O\left(\frac{b}{l_{d}}^{2}\right) \Big] 
    \approx \frac{\dot{R}_{x}}{32 D_{c}^{2}} \int_{0}^{t-1}  \frac{1}{(t-t')} = \frac{\dot{R}_{x}}{32 D_{c}^{2}} \ln(t).
\end{aligned}
\end{equation}
Thus, we have an explicit $t$ dependence in (\ref{eq:app_integrals}), a signature of history build-up of the chemicals due to (\ref{eq:curr}).  

%%%=======================================
\subsection{Symmetric trimers - ABA, BAB}
\label{app:dim_red_ABA}

For the ABA and BAB configurations, to solve for $\phi$ (bond angle of the trimer), we note that the cosine rule,  
\begin{align}
    (\Delta y_{13})^{2} = 8b^{2} \Big( 1 - \cos(\phi) \Big),
\label{eq:app_cos_rule}
\end{align}
implies the symmetry of the trimer problem means that the spring forces $F_{12} = -F_{21}$, $F_{23} = -F_{32}$, and $|F_{12}| = |F_{32}|$. We thus have the condition: $\frac{d(\Delta y_{13})}{dt} = 2v_{s} \sin (\theta_{1})$. Differentiating (\ref{eq:app_cos_rule}), and applying (\ref{eq:app_delta_y}) we arrive at (\ref{eq:phi_aba}). 
The symmetry of the trails (see Fig.~(\ref{fig12_m}c-d)), 
further enables us to reduce the effective dimensionality. 
%Using Eq. \ref{app:dim_red_mono}, 
We consider the $\theta_{1}$ equations for the ABA system:
\begin{align}
    \frac{d \theta_1}{dt} = \chi_{r}\Big (\cos(\theta_{1}){J}_{y_{1}} - \sin(\theta_{1}){J}_{x_{1}}\Big),\quad  {J}_{y_{1}} = \int dt^{'} \Big[ -\frac{\Delta y_{12}(t,t')}{\mathcal{D}^{2}(t,t')}\mathcal{F}^{(y)}_{12}%(t,t') 
    + \frac{\Delta y_{13}(t,t')}{\mathcal{D}^{2}(t,t')}\mathcal{F}^{(y)}_{13}%(t,t')
    \Big],
\label{eq:app_ABA_theta1}
\end{align}
% with 
where we have inserted entries from (\ref{eq:trinary_matrix}). We can show that ${J}_{y_{1}} = -{J}_{y_{3}}$ by noting the following relations (analogous to (\ref{eq:app_delta_y})), 
$\Delta y_{12} = -\Delta y_{32}, \quad \Delta y_{13} = -\Delta y_{31}
$ 
along with ${J}_{x_{1}} = {J}_{x_{3}}$ implies that 
$    \frac{d \theta_{1}} {dt} = -\frac{d \theta_{3}}{dt}
$, as is also evident in Fig.~(\ref{fig7}) and SI Video 8 \cite{siText}. Thus, at the fixed point orientations $\theta_{i}^{*}$, we have a stably propelling trimer governed by
  $  \frac{d R_x}{dt} = v_{s} \left( 2\cos(\theta_{1}) + \cos(\theta_{2}) 
    \right)$. Thus, we obtain equations (\ref{eq:ABA_red_fp}) and (\ref{eq:BAB_red_fp}). 

% \section{Trimers with AB(+): Results}
% \label{app:trimer}

% The results for ABA(+), BAB(+) trimers, are displayed here in Fig.~(\ref{fig14_t}). ABA(+) reproduces a chemo-repulsive trimer, BAB(+) reproduces a chemo-attractive trimer. Compare this with results of Fig.(\ref{fig7}).

\end{document}